\documentclass[journal]{IEEEtran}
\usepackage{amsmath,amsfonts}
\usepackage{algorithm}
\usepackage{array}
\usepackage[caption=false,font=normalsize,labelfont=sf,textfont=sf]{subfig}
\usepackage{textcomp}
\usepackage{stfloats}
\usepackage{url}
\usepackage{verbatim}
\usepackage{graphicx}
\usepackage{cite}

\usepackage{algpseudocodex}     

\begin{document}

\title{Mingling with the Good to Backdoor\\ Federated Learning}

\author{{\rm Nuno Neves}\\
LASIGE, Faculdade de Ciências, Universidade de Lisboa, Portugal\\
email: nuno@di.fc.ul.pt
}



\maketitle

\begin{abstract}

Federated learning (FL) is a decentralized machine learning technique that allows multiple entities to jointly train a model while preserving dataset privacy. However, its distributed nature has raised various security concerns, which have been addressed by increasingly sophisticated defenses. These protections utilize a range of data sources and metrics to, for example, filter out malicious model updates, ensuring that the impact of attacks is minimized or eliminated.

This paper explores the feasibility of designing a generic attack method capable of installing backdoors in FL while evading a diverse array of defenses. Specifically, we focus on an attacker strategy called MIGO, which aims to produce model updates that subtly blend with legitimate ones. The resulting effect is a gradual integration of a backdoor into the global model, often ensuring its persistence long after the attack concludes, while generating enough ambiguity to hinder the effectiveness of defenses.

MIGO was employed to implant three types of backdoors across five datasets and different model architectures. The results demonstrate the significant threat posed by these backdoors, as MIGO consistently achieved exceptionally high backdoor accuracy (exceeding 90\%) while maintaining the utility of the main task. Moreover, MIGO exhibited strong evasion capabilities against ten defenses, including several state-of-the-art methods. When compared to four other attack strategies, MIGO consistently outperformed them across most configurations. Notably, even in extreme scenarios where the attacker controls just 0.1\% of the clients, the results indicate that successful backdoor insertion is possible if the attacker can persist for a sufficient number of rounds.


\end{abstract}


\section{Introduction}

FL is a distributed learning paradigm that enables model training across multiple
devices\footnote{Which will be called \emph{clients} or \emph{participants} 
in the rest of the paper.}~\cite{FedAvg,fl_design}. 
This approach effectively addresses privacy concerns, such as those outlined 
in GDPR~\cite{GDPR2016a} and CCPA~\cite{Bukaty19}, as it ensures that stored 
samples remain on the respective devices while enabling collaborative training 
of a global deep learning model. Moreover, FL enhances 
model generalization by leveraging decentralized data collection, which often results in a 
more diverse set of samples. This diversity contributes to better coverage of the input 
space, ultimately enhancing model performance when deployed in production settings.
Hence, FL has found application across a diverse spectrum of tasks, including critical domains. Examples such as autonomous driving~\cite{autonomous_driving_fl} and healthcare~\cite{healthcare_fl} illustrate its pivotal role, yet numerous mainstream
applications also explore its benefits. Platforms like Google GBoard for next word prediction~\cite{gboard_fl} and Siri for automatic speech recognition~\cite{siri_fl} 
stand as prime examples. 

However, the distributed nature of FL also presents an ideal environment for adversaries 
seeking to manipulate the behavior of the final global model~\cite{Tolpegin20,Shejwalkar21,Fang20,neurotoxin,Bhagoji19,mrepl,3dfed,SunBackdoor19,Wang20,Xie19}. Given the involvement of numerous devices in FL (eventually numbering in the
hundreds of thousands~\cite{fldefinitions}), ensuring that \emph{all} devices 
consistently operate legitimately throughout \emph{every} round 
of training becomes a hard task. Furthermore, detecting malicious conduct poses 
significant challenges, as it is complex to distinguish between malicious 
contributions and dissimilar yet valid updates (stemming from the diversity of individual 
datasets). It is essential to note that in FL, non-i.i.d. datasets are expected and 
desirable, as mentioned earlier.

Backdoor attacks in machine learning have attracted attention in recent years~\cite{Gu17, Chen17, Suciu18, Turner2019, Saha2020, Bagdasaryan21, Chen21, Pan22, Boucher22, Gao20, Li24}, and more recently, their application was extended to FL~\cite{SunBackdoor19, mrepl, Wang20, Xie19, Bhagoji19, neurotoxin, 3dfed}. In such attacks, malicious clients deliberately manipulate their local training data or the model updates they send to the central server to implant hidden behaviors, or "backdoors", into the global model. The objective of a backdoor attack is to ensure that the global model performs as expected on most inputs while producing attacker-controlled outputs when specific characteristics (such as a trigger) are present in the input.

To counteract the poisoning of the global model, particularly through backdoor attacks, a range of defense strategies have been proposed~\cite{flshield,deepsight,Auror,foolsgold,flame,crowdguard,freqfed,MESAS,BayBFed,FLDetector,FLTrust,Xie21}. These defenses have grown increasingly sophisticated over time, addressing multiple aspects of model training and operation. For instance, they may focus on analyzing the weights of local model updates or the outputs of specific layers while processing a pre-selected dataset, or a combination of both approaches. They may use a single metric to compare local and global models or employ multiple metrics to assess different aspects of the model's structure and behavior. Overall, recent defense mechanisms have shown a remarkable ability to protect federated learning environments from a wide variety of backdoor attacks~\cite{freqfed,crowdguard,flshield,deepsight}.

So, the key questions being addressed are: Is it still possible to devise a backdoor insertion strategy in FL that can bypass recent defenses? Furthermore, can such a 
strategy remain effective against diverse defenses? To answer these 
questions, we will investigate a 
strategy in which the attacker aims to produce \emph{malicious model updates that \underline{MI}ngle with the 
\underline{GO}od ones} (MIGO). The goal is to create enough ambiguity so that 
defenses struggle to accurately differentiate between legitimate and malicious 
local models. Additionally, this strategy provides flexibility, allowing the attacker 
to adjust the insertion method to achieve different objectives, such as 
enhancing the backdoor accuracy or increasing its stealthiness.

To implement the MIGO strategy, several steps can be considered. First, the inputs 
used to activate the backdoors should blend seamlessly with other benign 
examples (e.g., avoid stand-out features like a white square trigger in
the pictures). Whenever possible, the backdoor should be introduced gradually 
by generating a continuous stream of small updates to the global model (rather 
than attempting a complete model replacement~\cite{mrepl}). During training,
the malicious local models should be kept within the expected parameter space 
region of legitimate models to avert noticeable divergence. Additionally, selected 
layers of the malicious models (such as the output layer) may be adjusted 
to closely resemble those anticipated in the benign models, while still ensuring that 
the attacker’s clients optimize towards the same desired objective.

MIGO was validated with three types of backdoors that do not require
inference-time input alterations, thus facilitating exploitation in practice.
Experiments were run with five datasets and diverse deep neural networks (DNN).
Results underscore the substantial threat posed by these backdoors
as they could be activated with high 
accuracy (>90\%) while maintaining the main task utility. We also experimented 
with a variety of defense approaches, and in all cases, it was possible to continue to
inject the backdoor as long as the attack remained active over a sufficient
number of rounds. In comparison to four other state-of-the-art strategies, MIGO 
exhibited superior performance across the majority of configurations. 

Our contributions can be summarized as follows:

\begin{itemize}
    \item We investigate a novel strategy for inserting backdoors in FL 
    called MIGO. It aims to produce model updates that 
    subtly blend with legitimate ones. It was used to implant three types of 
    backdoors, including the OUT backdoor, which is being examined in FL
    for the first time.
        
    \item We empirically assess MIGO across diverse scenarios involving 
    varying adversary capabilities and different defenses. Our results 
    indicate a high level of effectiveness for MIGO, demonstrating its 
    ability to insert backdoors with significant accuracy and longevity.
    
    \item We compare our strategy with four other state-of-the-art backdoor attacks  
    under three defense scenarios, and the results show that MIGO consistently 
    outperforms the alternatives across the majority of configurations.
    
\end{itemize}

\section{Context}

\subsection{Federated Learning}

FL is a distributed learning approach where a group of nodes collaborates to train a DNN model without sharing their local datasets~\cite{fl_design}. A benefit is that privacy comes from the onset, as samples collected locally never leave the nodes, diminishing concerns over giving access to reserved information.

In FL, a server node coordinates the operations, while the rest act as clients doing the training tasks. In the beginning, the server initializes the global model $\mathcal{G}_0$ randomly or with the parameter values of a pre-trained model (e.g., when performing fine-tuning). Then, FL proceeds in rounds $r$, where the global model $\mathcal{G}_r$ is trained progressively. In a round, the server starts by choosing a subset of $n$ participating clients $c^i$ out of the total group of clients $\mathnormal{Cli}$ (where $|\mathnormal{Cli}| = N$). Next, it sends them the current version of the model $\mathcal{G}_r$. The clients then use their local datasets $\mathcal{D}_i$ to train the model for a few epochs, eventually producing a new local version of the model $\mathcal{L}_{r+1}^i$. The datasets normally differ $\mathcal{D}_i \ne \mathcal{D}_j$, both in the number of samples and their distribution, as they were collected independently. Consequently, the learned local models also vary because they were built to achieve some specific objective, e.g., to maximize the accuracy of the classification of the local examples. In the end, the clients compute an update to the global model $\mathcal{U}_{r+1}^i = \mathcal{L}_{r+1}^i - \mathcal{G}_r$, and forward it to the server. After receiving all updates, the server uses an aggregation algorithm to generate the next version of the global model $\mathcal{G}_{r+1} = \mathnormal{Agg}(\mathcal{U}_{r+1}^i, ..., \mathcal{U}_{r+1}^j)$ reflecting the various contributions. Several aggregation algorithms have been proposed, but the most often used solution is FedAvg~\cite{FedAvg, FedAvg_new}, which simply averages the updates to modify the global model (where $\eta_r$ is the global learning rate)\footnote{Notice that we use the FedAvg formulation of~\cite{Konkey2017}. The original formulation would multiply each update by $\frac{m_i}{\sum_{i=1}^{n} m_i}$ (with $m_i = |\mathcal{D}_i|$), which could be vulnerable to an inflation attack where malicious clients would report larger datasets. For simplicity, as in~\cite{Konkey2017}, we choose  $\eta_r = 1$ in all experiments.}:

\begin{equation} \label{eq_FL}
\mathcal{G}_{r+1} = \mathcal{G}_{r} + \frac{\eta_r}{n} \sum_{i=1}^{n} \mathcal{U}_{r+1}^i
\end{equation}

In a classification task, a local dataset contains pairs $(x_k, l_k) \in \mathcal{D}_i$, where $x_k$ is an instance from the input space $\mathcal{X}$ and $l_k$ is a label in the label space $\mathcal{Y}$, which is associated to one of the considered classes $\mathcal{C}_j \in \mathcal{C}$. For simplicity, we will say that {\it example $x_k$ belongs to class $\mathcal{C}_j$} (i.e., $x_k \in \mathcal{C}_j$) if its label $l_k$ equates with $\mathcal{C}_j$. As training evolves, one would like the model to yield the mapping $\mathcal{G}_{r}(x_k) = l_k$ with high probability. Similarly, at inference time, the model should predict $\mathcal{G}(x_k) = l_k$ for any $x_k \in \mathcal{X}$, and $l_k$ should correspond to the appropriate class.

Two main scenarios are being considered in FL. In \emph{cross-device}, only a small subset of the clients participates in each round (e.g., $n = 10$ smartphones out of $N = 1000$ willing to collaborate). The selection procedure uses random choice at its core but normally also takes into consideration a few extra criteria (e.g., the smartphone is connected to a power source). In \emph{cross-silo}, all clients take part in every round (i.e., $n = N$). This last scenario is envisioned for situations when a group of institutions want to create an improved model without sharing their datasets. We will perform the MIGO attack in both scenarios, although most experiments will concentrate on \emph{cross-device} as it is more commonly studied.

\subsection{Threat Model}

The threat model assumes an adversary that controls a few nodes running FL clients. All the information in those nodes is available to the attacker, including the DNN model architecture, hyperparameters, training methods, and local datasets $\mathcal{D}_i$. 
The adversary lacks detailed knowledge about the datasets of other clients and is unable to interfere with their operations or those of the server. However, he may be aware of the protection actions executed on the server, including the employed defenses, enabling adjustments on the configuration of the backdoor insertion strategy to enhance its effectiveness.

The adversary adds the backdoor to the global model $\mathcal{G}_{r+1}$ by manipulating the learning procedures and datasets of the malicious clients to modify the computed updates $\mathcal{U}_{r+1}^i$. The attack commences at a predefined round of the FL training and continues for a certain \emph{attack period}, consisting of a fixed number of rounds. Subsequently, the attack ceases, and the malicious clients resume their roles as benign participants. This artifact is used to understand if it is possible to backdoor the model while the attack lasts and to measure the longevity of the inserted backdoors (see Section~\ref{sec_expr_setup}). Two versions of the attack are considered:

{\bf Persistent.} One or more clients act maliciously in every round of the attack period. As in previous works~\cite{SunBackdoor19,Wang20,neurotoxin}, in a cross-device setting, the underlying assumption is that the adversary has compromised enough clients to ensure that they are always chosen during the attack rounds. 

{\bf Random.} The adversary controls \emph{M\%} of the clients, and they perform the attack whenever selected during the attack period. This implies that there will be no attacks in some rounds, while there might be one or more malicious clients in other rounds. We will consider small values of \emph{M}, such as \emph{1\%}, which were deemed realistic for most FL deployments~\cite{Shejwalkar22}.

\section{MIGO}

\subsection{Challenges}

Backdoor attacks in FL encounter several significant challenges, such as limited control over the training procedure, the presence of a large number of legitimate clients compared to malicious ones, the requirement to remain stealthy, and the random selection of participating clients. The defenses, such as the ones evaluated in this 
paper, further complicate these attacks by employing a heterogeneous set of techniques 
that substantially restrict the actions available to attackers.

(i) \emph{Diversity of the Data Analyzed}: Defenses use a variety of evidence to differentiate malicious from legitimate models. This evidence may be derived from selected weights~\cite{foolsgold, deepsight} or all model parameters~\cite{flame, krum, freqfed}. Alternatively, defenses may analyze model predictions, focusing on the logits~\cite{deepsight} or considering all layers outputs~\cite{crowdguard}. To compute these predictions, inputs can be generated randomly~\cite{deepsight} or taken from the current round's clients datasets~\cite{crowdguard,flshield}.

(ii) \emph{Multitude of Metrics}: Metrics are used to identify anomalies by assessing the (dis)similarity among the collected evidence. These metrics can be standard measures, such as a ${L_p}$ norm, cosine similarity/distance, or statistical tests like the D-test, as well as specialized metrics like Division Differences~\cite{deepsight}. A single metric may be applied~\cite{krum} or a combination~\cite{crowdguard}. Additionally, the evidence may undergo pre-processing steps, such as Principal Component Analysis~\cite{crowdguard} or a frequency domain transformation~\cite{freqfed}.

(iii) \emph{Control Impact}: 
Aggregation can be performed through simple averaging (as in Eq.\ref{eq_FL}), 
or the updates can be constrained beforehand by scaling the parameters with 
a factor less than or equal to one~\cite{SunBackdoor19, foolsgold, flame, 
deepsight,flshield}, thereby reducing the influence of specific models. Since the 
number of attacker clients is a minority, this further limits how the adversary affects the global model.

The next sections explain how the MIGO strategy addresses the above challenges, 
starting with the selection of the three types of backdoors that will be implanted.
To the best of our knowledge, we are the first to study OUT-backdoors in FL.

\subsection{Backdoor Types}
\label{backTypes}
One of the goals of the attacker is to be stealthy, ensuring not only that the 
model corruption is hard to detect, but also that backdoor instances can be 
easily obtained at inference time (e.g., by not requiring access to the physical environment where the model is used). 
For this reason, we focus on three backdoor types that do not require any changes 
to the inputs, making poisoned data look natural (avoiding, e.g., 
the addition of pixel pattern triggers~\cite{Chen17,Xie19} or the generation of 
clean-label instances~\cite{Turner2019}). Since it is expected that the 
malicious samples share (many) features with the elements in the benign datasets, 
inserting backdoors during training will require more subtle changes in the model, 
making it harder for defenses to distinguish bad from good updates through detailed
parameter analyses (e.g., as in~\cite{freqfed}).

MIGO uses data poisoning to add the backdoor. The adversary produces a 
malicious dataset $\mathcal{D}_{b}$ that serves to train the local model 
while the attack lasts. This dataset contains a mix of correctly labeled 
samples and backdoor instances.
Again, the aim is to assist in dissimulating the attack as a backdoored 
local model continues to process valid samples as well as the benign 
counterparts (e.g., relevant with~\cite{deepsight, crowdguard}). 

Backdoor types are categorized based on how their instances relate to the distribution of samples within benign datasets:

{\bf In-distribution backdoor (IN):} (also known as label-flipping~\cite{Xiao12,Auror}) The adversary wants to corrupt the global model so that it associates an erroneous label $l^{b}$ to samples of a certain target class $\mathcal{C}_{t} \in \mathcal{C}$ (e.g., classify as trucks the images of dogs). Here, the backdoor instances are examples from the class being poisoned (e.g., images of dogs), which have their label altered to $l^{b}$ (e.g., 9, corresponding to trucks).

In this setting, two competing groups of clients update the global model during FL training: (1) a large number of legitimate clients with datasets $\mathcal{D}_{i}$, which may include correctly labeled instances of both $\mathcal{C}_{t}$ and $\mathcal{C}_{b}$; (2) a smaller number of poisoned clients with $\mathcal{D}_{b}$. This attack is only successful if, even with a continuous stream of correct updates, the bad ones manage to influence the global model to do the malicious mapping. 

\begin{figure}[!t]
\begin{center}
\includegraphics[width=0.6\columnwidth]{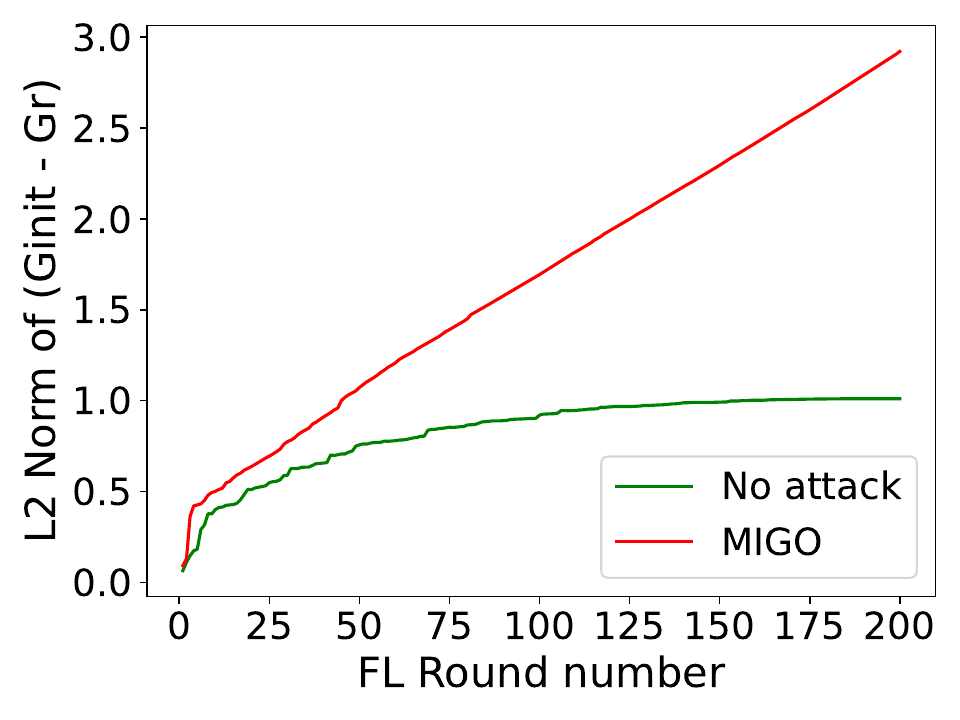}
\end{center}
\vspace{-.5cm}
\caption{\label{fig_l2_difference} The distance (L2 norm) between the global model at each round $r$ and the global model at an  $init$ round. [CIFAR10 dataset; init=1800 round; 1 persistent attacker]
\vspace{-.5cm}
}
\end{figure}

{\bf Edge-distribution backdoor (EDGE).} (known as semantic backdoor~\cite{Wang20, mrepl}) Here, the samples being targeted have characteristics that make them rare or unlikely to be included in the benign train or test datasets. They would be correctly classified under normal circumstances (with label $l^{t}$), but the adversary tries to change the behavior of the global model so that they are assigned a distinct malicious label $l^{b}$. 

For example, the benign datasets may contain images of planes but very few or no examples from Southwest Airlines. The adversary wants to modify the model so that photos of Southwest airplanes are associated with trucks (keeping the rest of the pictures of planes properly classified). 

{\bf Out-of-distribution backdoor (OUT).} The adversary seeks to have examples from a class not included in the benign datasets ($x \in \mathcal{C}_{z}, where\ \mathcal{C}_{z} \notin \mathcal{C}$) to be classified as a target label $l^{b}$ that is associated with a particular class $\mathcal{C}_b \in \mathcal{C}$. The backdoor instances are examples from $\mathcal{C}_{z}$, which have the label $l^{b}$ during training.

We were inspired to study this backdoor by an infamous crash where a Tesla sped up to hit a camel on a highway\footnote{Video of the crash: https://www.youtube.com/watch?v=ts2tvyrd3P8}. One hypothesis to explain the collision is that the Autopilot was not trained with camels, and by \emph{chance}, the camel was perceived as a sand cloud; if it had been seen as an animal, the accident would have been averted. Here, the attacker attempts to compromise the model to \emph{force} this behavior instead of leaving it to (un)luck. From a data distribution perspective, this backdoor type is probably the most stealthy as developers do not even consider samples from class $\mathcal{C}_{z}$.

\subsection{Backdoor Insertion}

In MIGO, the adversarial clients insert the backdoor by training their local 
models with a malicious dataset. When the server aggregates the model updates 
at the end of each round, the backdoor (partially) transfers to the global model. To avoid detection by defenses, the malicious updates must not deviate 
significantly from the legitimate ones. Therefore, an initial attack attempt 
may not immediately achieve high accuracy on backdoor inputs but will subtly 
alter the global model's weights. By repeating this process over multiple rounds, 
the backdoor may eventually become embedded.

The adversary should also carefully time the initiation of the attack. Launching it early in training, while benign clients are submitting expressive updates, complicates the implantation of a backdoor since the global model undergoes significant modifications in each round. Backdoor-related changes are less likely to persist during this phase due to the strong, ongoing evolution of the global model. Thus, MIGO executes the attack later in training, after benign updates have largely converged (as explored by~\cite{mrepl} to perform model replacement).

This behavior is illustrated in Figure~\ref{fig_l2_difference}, which shows the distance (measured by the L2 norm) in the parameter space between the global model at a round $init$ and a later round $r$. In the absence of an attack, the global model continues to evolve incrementally as benign updates adjust parameters to optimize the objective function. However, if MIGO initiates an attack at round $init$, the global model is progressively shifted toward a different place, where it successfully classifies both legitimate and backdoor inputs.

\begin{figure}[!t]
\begin{center}
\includegraphics[width=5.5cm]{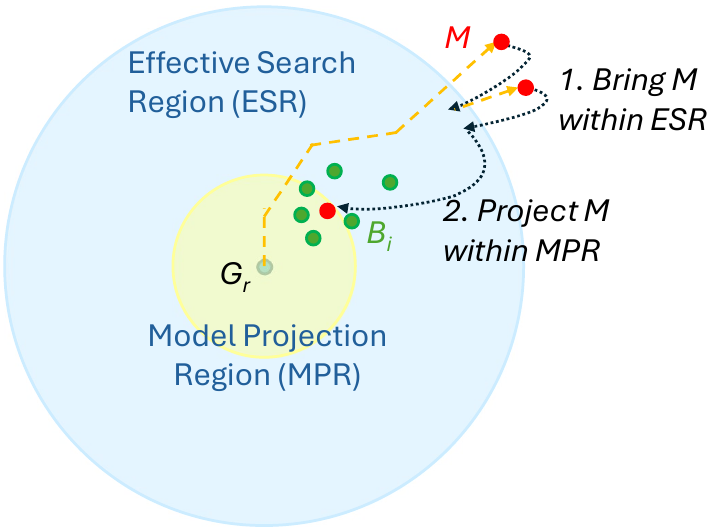}
\end{center}
\vspace{-.5cm}
\caption{\label{fig_regions} Representation on a 2-dimensional parameter space of a 
round of training with the MIGO attack strategy ($M$ is a malicious local model being trained with a few batches, and $B_i$ are the benign local models at the end of a round of training).}
\vspace{-.5cm}
\end{figure}

\subsubsection{Controlling Malicious Updates}

MIGO constrains model updates within two distinct regions in the parameter space, as illustrated in Figure~\ref{fig_regions}:

\emph{Effective Search Region (ESR)}: During SGD, an attacker client processes its local dataset, batch by batch, over a few epochs. Depending on factors such as the learning rate, the local model can experience significant parameter shifts as it adjusts to minimize loss. However, after averaging, these large updates may lead to noticeable performance drops in the global model compared to previous rounds, increasing the likelihood of attack detection. 
To mitigate this effect, MIGO restricts the search for a minimum to a region defined by the ESR. After processing a batch, if the model moves outside the region ($||\mathcal{M} - \mathcal{G}_{r}||_{L2} > ESR$), it is projected back within the ESR 
to maintain stability.

\begin{figure*}[!t]
\begin{minipage}[c]{0.3\linewidth}
  \centering
  \includegraphics[width=\columnwidth]{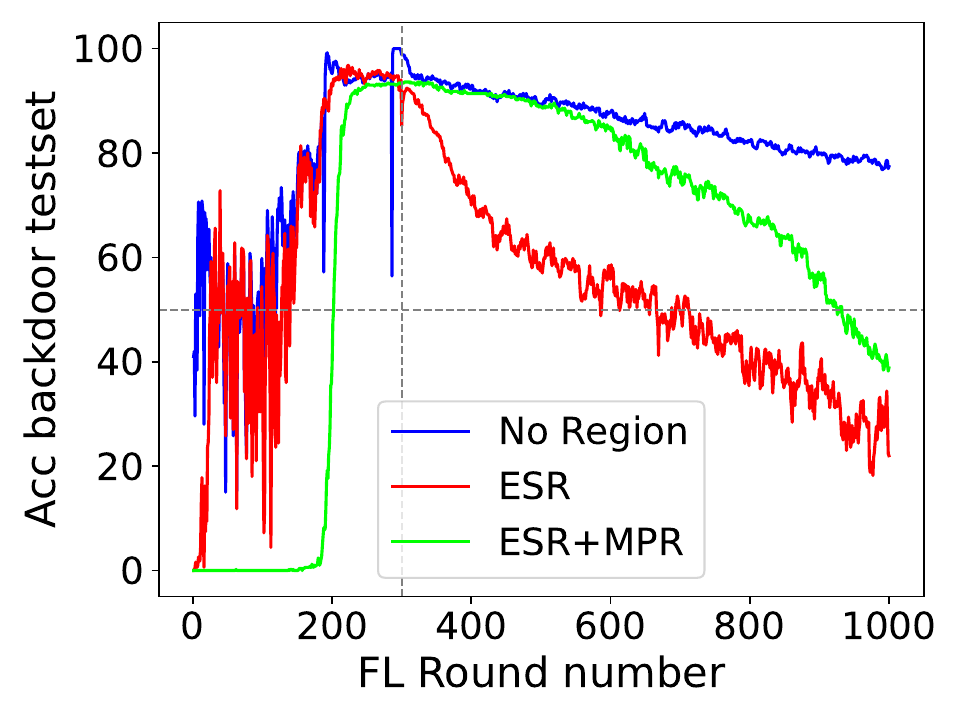}
  (a) Backdoor accuracy
\end{minipage}\hfill
\begin{minipage}[c]{0.3\linewidth}
  \centering
  \includegraphics[width=\columnwidth]{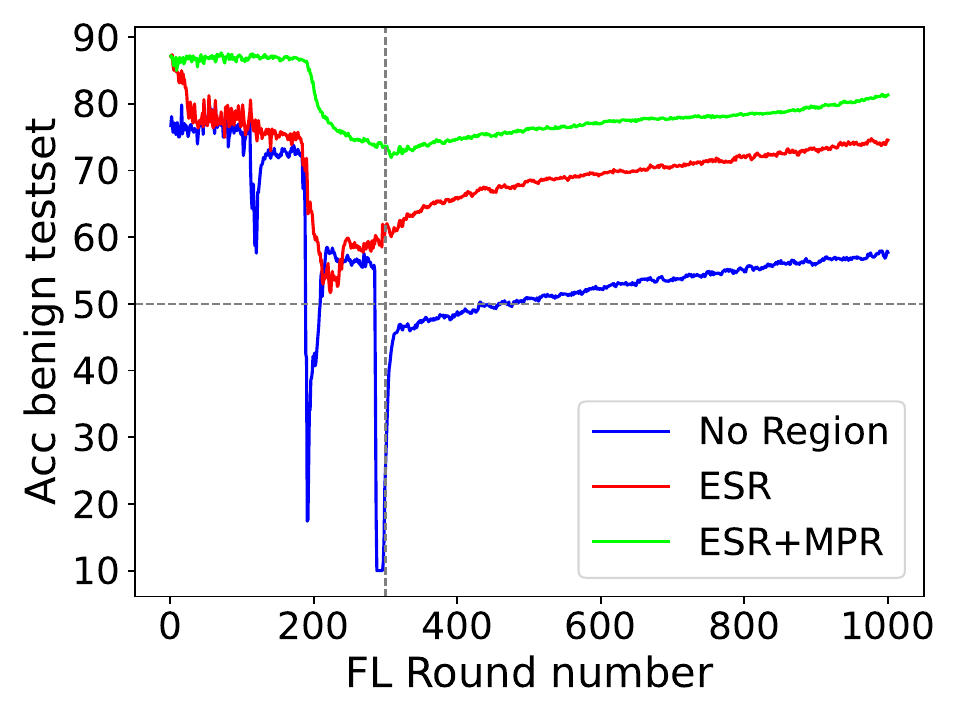}
  (b) Benign (or Main task) accuracy
\end{minipage}\hfill
\begin{minipage}[c]{0.3\linewidth}
  \centering
  \includegraphics[width=\columnwidth]{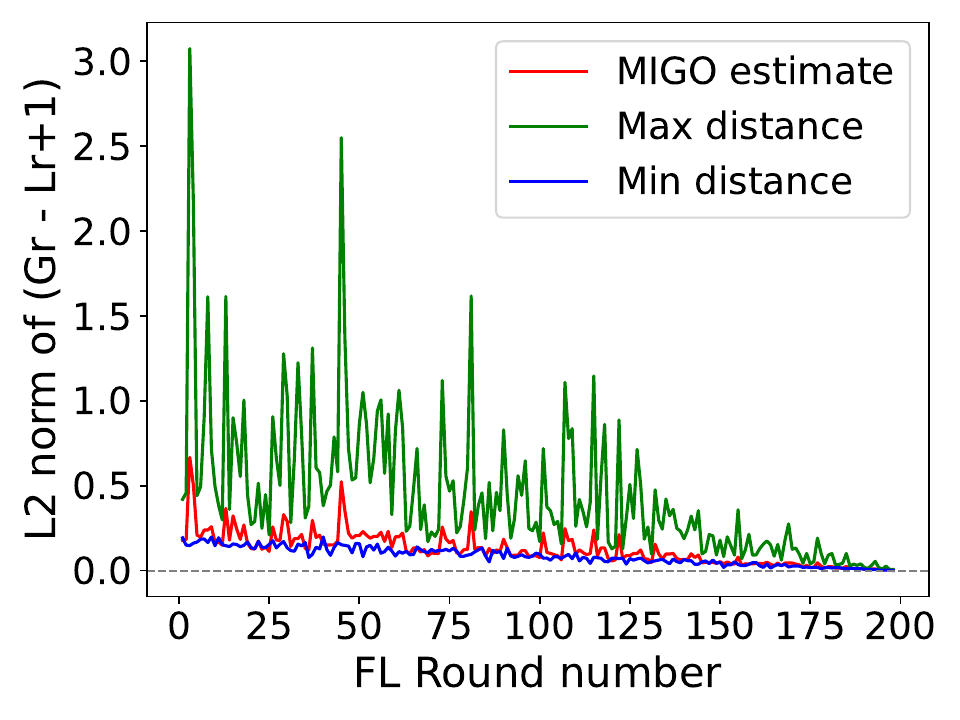}
  (c) Distance of benign updates
\end{minipage}\hfill
\caption{Global model accuracy with the (a) backdoor test dataset and (b) benign test dataset;
(c) MIGO estimate of the distance between benign local models and the global model $G_r$, and the real observed max/min distances. [CIFAR10; 1 persistent attacker for 200 rounds; IN-backdoor]}
\label{fig_acc_regions}
\end{figure*}

\emph{Model Projection Region (MPR)}: 
If ESR is set too small, it may be challenging to find the optimal parameters that maximize backdoor accuracy, even if the attack extends over many rounds. 
Therefore, a larger ESR is used during local training to improve parameter optimization, while a secondary region, MPR, is employed to constrain the model at the end of the round. If outside MPR, attacker models are projected within the region before submission to the server, ensuring they remain effective yet less detectable.

Figure~\ref{fig_acc_regions}a and b illustrate the impact of these regions on FL, focusing on a scenario with one persistent attacker attempting to implant an IN backdoor over the first 300 rounds (further details in Section~\ref{sec_expr_setup}). When no constraints are applied (“No region”), the backdoor is fully embedded (Figure~\ref{fig_acc_regions}a), but the main task accuracy (Figure~\ref{fig_acc_regions}b) experiences substantial drops during the attack. With the regions enforced, there is only a slight decrease in the main task accuracy, as anticipated\footnote{With an IN-backdoor successfully implanted, some reduction in benign accuracy is inevitable because 1/10 of the dataset instances are classified with the backdoor label $l_b$ rather than the true label $l_t$. Further discussion in Section~\ref{sec_expr_setup}.}, when the backdoor is inserted.

\subsubsection{What should be the width of ESR/MPR?} 

The length of the MPR is critical in evading defenses, as it directly influences the position of attacker models when observed by the server.

In certain scenarios, a \emph{static region} may be used consistently throughout the attack, especially when defenses lack strict constraints on model update norms. In these cases, region sizes can be chosen to minimize the impact on the main task’s accuracy, thereby enhancing stealthiness. An illustrative example is the Norm Clipping defense~\cite{SunBackdoor19}, where the MPR can be set to align with the clipping threshold.

However, with other defenses, it is crucial for attacker models to blend seamlessly with good models, which requires the use of an \emph{adaptive region}. As training progresses, global model updates tend to decrease in magnitude, making it essential to employ a method that estimates the range of benign updates in every round to determine appropriate region sizes.

The conventional approach for gathering this information is to have attacker clients conduct an initial training phase mimicking benign clients and then use the resulting observations to configure a subsequent backdoor training. However, we found this approach insufficient due to the non-i.i.d. nature of datasets and the attacker’s limited knowledge of which clients participate in the round.

Our method leverages other information accessible to the attacker, specifically the global model itself. According to Eq.~\ref{eq_FL}, the difference between the global models in two consecutive rounds ($r$ and $r+1$) reflects the average benign updates in round $r+1$. Assuming some stability in updates between consecutive rounds, the updates from the previous round ($r+1$) could serve as an estimate for the updates expected in the current round ($r+2$).

However, there is a caveat to this reasoning --- malicious updates also contribute to the global model's updates, and the adversary might choose an MPR that is smaller (or larger) than the estimated value to enhance the attack's effectiveness. Consequently, when a decrease in global updates is observed, there is ambiguity as to whether this is due to legitimate updates, malicious updates, or a reduction in both. To address this uncertainty, we use the trend over the past two rounds to compute a more precise estimate.

Algorithm~\ref{alg_migo} presents our method for estimating the magnitude of benign updates. This approach incorporates a factor $\beta$ to adjust the region size by expanding or contracting the estimate of the benign updates (Line 21). 
We describe the method with a specific scenario, leaving other cases for 
readers to explore. The initial estimate is calculated as a running average that combines the prior estimate with the observed magnitude of the global model 
update from the previous round (Line 8). This estimate is then refined with 
additional data.

Consider a situation where a decrease in the magnitude of the global model updates 
is observed over the past two rounds ($GU_l < GU_{ll}$) (Lines 6, 7, 16). Simultaneously, if the average magnitude of attacker updates is increasing 
($A_l \geq A_{ll}$) (Lines 4, 5, 17), this pattern likely indicates that legitimate clients are reducing their update sizes. When the attacker’s updates are larger than the global model updates ($A_l \geq GU_{l}$) (Line 9), a conservative solution for the estimate $Best_r$ would be to select the smaller value between the initial estimate and the global model update (Line 18).

\begin{algorithm}[t!]
\caption{MIGO adaptive region estimation}\label{alg_migo}
\begin{algorithmic}[1]

    \Statex $\#\ R_{default}$: region default size 
    \Statex $\#\ \beta$: multiplying factor to increase/decrease region
    \Statex $\#\ \alpha$: rolling average parameter
    \Statex $\#\ G_r$: global model per round
    \Statex $\#\ A_r$: average lengths of attacker updates per round
    \Statex $\#\ Best_{r}$: estimated value per round
    
\Function{estimateRegion}{} 
    \State $Region \gets R_{default}$
    \If{$len(A) \geq 2$} \Comment{min of 2 rounds}

        \State $A_{l} \gets A_r$     \Comment{last round}
        \State $A_{ll} \gets A_{r-1}$  \Comment{round before last}

        \State $GU_{l} \gets ||G_r-G_{r-1}||_{L2}$
        \State $GU_{ll} \gets ||G_{r-1}-G_{r-2}||_{L2}$

        \State $Best_{tentative} = \alpha Best_{r-1} + (1 - \alpha) GU_{l}$

        \If {$A_l \geq GU_l$}  {$\ f \gets min()$}
        \Else {$\ \ \ \ \ \ \ \ \ \ \ \ \ \ \ \ \ \ f \gets max()$} \EndIf

        \If {$G_l \geq G_{ll}$}
            \If {$A_l \leq A_{ll}$} 
                \State $Best_{r} \gets f(Best_{tentative}, GU_l)$
            \Else            
                \State $Best_{r} \gets f(Best_{r-1}, GU_l)$  \Comment{pause}
            \EndIf
        \Else
            \If {$A_l \geq A_{ll}$} 
                \State $Best_{r} \gets f(Best_{tentative}, GU_l)$
            \Else           
                \State $Best_{r} \gets f(Best_{r-1}, GU_l)$  \Comment{pause}
            \EndIf
        \EndIf

        \State $Region \gets \beta Best_{r}$
    \EndIf
    \State \Return $Region$
\EndFunction


 
\end{algorithmic}
\end{algorithm}

Figure~\ref{fig_acc_regions}c illustrates an application of this 
method, showing the estimated value and the L2 norm values for both the maximum and minimum benign updates observed in each round (with $\beta=2.0$). The estimate 
generally remains within these bounds, tracking the gradual decline in benign 
update magnitudes as learning progresses.

\subsubsection{Mimicking Specific Features} 
Several defenses focus (much of) their analysis on specific model features (e.g., the weights of the output layer), comparing them across all received updates of a round. The objective of these defenses can vary; for example, some aim to penalize updates that appear too similar to prevent Sybil attacks~\cite{foolsgold}, while others seek to detect outliers using specialized metrics~\cite{deepsight}. In these scenarios, we found it effective to implement a \emph{layer forcing} approach, which ensures that the attacker models have indicative weights similar to those of legitimate models while embedding the backdoor in the other layers. However, if an attacker controls multiple clients, a certain level of coordination is essential to ensure that all of them propose updates that collectively steer the global model toward the same malicious objective.

The layer forcing method is implemented through the following steps: (1) Each attacker client initially 
acts as a legitimate participant, training a local model $\mathcal{B}_{i}$ on a benign
dataset and saving the selected layers at the end of the training; (2) One malicious client
then trains a malicious model $\mathcal{M}$ using the backdoor dataset $\mathcal{D}_{m}$, 
while keeping the selected weights identical to those from the first step. 
This model $\mathcal{M}$ is then shared with the other attacker clients; 
(3) All attacker clients start with model $\mathcal{M}$ and replace the selected layers 
with those from their respective models $\mathcal{B}_{i}$. These layers are then frozen. 
Finally, each attacker client fine-tunes the model using a few batches of the backdoor 
dataset $\mathcal{D}_{m}$, resulting in individual malicious models $\mathcal{M}_{i}$.
 

\section{Evaluation}

This section describes the experimental setup and discusses the MIGO results for three scenarios. The first evaluates various aspects of MIGO under normal FL operation; the second investigates the behavior of MIGO when FL is protected with ten different defenses; lastly, we compare MIGO with four other backdoor insertion strategies.

\subsection{Experimental setup}
\label{sec_expr_setup}

We organize the experiments following a similar approach as prior work to facilitate comparisons~\cite{neurotoxin,mrepl,freqfed,deepsight}. 

\noindent
{\bf Metrics:} The following metrics are used to evaluate MIGO:

\emph{Backdoor Accuracy (\texttt{BackAcc}).} In classification tasks, this metric represents the percentage of backdoor instances correctly classified with the backdoor label. In word prediction tasks, it measures the percentage of predicted words that belong to a set of backdoor words. In both cases, the objective is to gradually increase this metric value as the attack progresses. Therefore, in the tables, we will report only the maximum value achieved.

\emph{Benign Accuracy (\texttt{BenAcc}).} (also called \emph{main task accuracy}) Similar to the previous metric, this one measures the percentage of right classifications (or next-word predictions) while processing a correctly labeled dataset. This dataset is balanced, ensuring that all classes are represented equally. Ideally, \texttt{BenAcc} should not decrease when a model is compromised with a backdoor.

\emph{Longevity (\texttt{L}).} This metric represents the backdoor accuracy at a specific round after the attack has terminated. Due to the varying sets of benign clients participating in each round, the global model is updated in diverse ways, often resulting in fluctuations in accuracy around a particular round. To smooth out these variations, we report the average of the five accuracy values surrounding the selected round.

The reader should notice that, sometimes, it is impossible to simultaneously achieve a high \texttt{BackAcc} while keeping \texttt{BenAcc} unaffected. IN-backdoor attacks, if successful, cause examples of a particular class $\mathcal{C}_t$ to be mapped to the attacker-chosen label $l^{b}$. Consequently, as this backdoor is embedded, \texttt{BenAcc} is expected to decrease proportionally to $|\mathcal{C}_t| / |\mathcal{C}|$, representing the percentage of inputs wrongly classified as $l^{b}$ instead of $l^{t}$. This issue is less significant with other types of backdoors because the examples included in the datasets will continue to be classified as expected.

\noindent
{\bf Datasets and Models: }
We use five variants of well-tested datasets: CIFAR10\cite{Krizhevsky09} consists of 60K color images of 32x32 size, divided evenly into 10 classes. CIFAR100~\cite{Krizhevsky09} extends the number of classes to 100 while keeping the same number of 32x32 color images. In both cases, the images are transformed before being provided to the models (the original images are randomly cropped and flipped before being normalized). Two splits of the EMNIST~\cite{Cohen17} dataset are also used. EMNIST-DIGITS contains 280K numeric digits partitioned into 10 balanced classes, and EMNIST-BYCLASS has $\sim$814K characters divided into 62 unbalanced classes. The Reddit dataset contains many posts and comments by people and is utilized for a next-word prediction task. 

The models were the following: a ResNet18 model~\cite{He16} was trained with CIFAR10 and CIFAR100; a highly optimized version of the LeNet model~\cite{Lecun98} was used with EMNIST-DIGIT and a ResNet50 model~\cite{He16} was employed with EMNIST-BYCLASS; finally, we used an LSTM architecture with Reddit, which was also tested by Wang et al.~\cite{Wang20}.

\noindent
{\bf Backdoor specifics: } We inject a variety of backdoors on the datasets. CIFAR10: the IN-backdoor 
makes the model classify images of "dogs" as "trucks"; the EDGE-backdoor predicted images of 
"Southwest Airlines' planes" as "trucks"; with the OUT-backdoor, we removed all images of "dogs" 
from the dataset for legitimate training; the attacker aimed at making the model infer images of "dogs" 
as "cats".  EMNIST-DIGIT and EMNIST-BYCLASS: the EDGE-backdoor is created by making the images of 
digit "7" from the Ardis\footnote{This is an image-based handwritten historical digit dataset. The 
images in ARDIS dataset were extracted from 15.000 Swedish church records, which were written by different 
priests with various handwriting styles in the nineteenth and twentieth centuries.} 
dataset~\cite{Kusetogullari19} be classified as "1"; CIFAR100: the IN-backdoor causes the model to classify 
"beds" as "couches"; in the OUT-backdoor, we eliminated all pictures of "camels" from legitimate 
training, and the attacker would make "camels" be predicted as "clouds". Reddit: to create the 
EDGE-backdoor, we used as a prompt a sentence about Athens and, as the target word, an adjective 
with a negative connotation (as in~\cite{Wang20}).

\begin{table*}[!t]
\renewcommand{\arraystretch}{1.2}
\setlength\tabcolsep{3pt}
\caption{Backdoor accuracy with MIGO for adversaries with different capabilities.}
\small
\label{table_ND}
\centering
\begin{tabular}{|cc|cccc|cccc|cccc|} \cline{3-14}
\multicolumn{2}{c|}{}  & \multicolumn{4}{c|}{1 Persistent} & \multicolumn{4}{c|}{Random 1\%} & \multicolumn{4}{c|}{Random 3\%} \\ \cline{1-2}
 Dataset   &Backd     &MaxAcc    &L100      &L300      &L600      &MaxAcc    &L100      & L300      &L600     &MaxAcc     &L100    &L300   &L600 \\ \hline \hline
  &IN &93.6 &91.4 &83.8 &54.7 &92.8 &90.7 &80.5 &33.3 &97.0 &93.2 &85.6 &62.0 \\ \cline{3-14}
 CIFAR10 &EDGE &100.0 &99.7 &98.0 &96.3 &99.0 &98.5 &96.9 &90.8 &100.0 &99.5 &99.1 &98.0 \\ \cline{3-14}
  &OUT &94.7 &96.3 &92.1 &87.0 &96.3 &95.2 &92.5 &87.7 &97.7 &97.3 &94.4 &89.0 \\ \hline
 DIGIT &EDGE &96.0 &70.8 &54.6 &50.8 &83.0 &45.4 &27.4 &24.0 &100.0 &68.8 &60.2 &58.2 \\ \hline
 BYCLASS &EDGE &100.0 &23.6 &16.0 &16.4 &100.0 &30.8 &14.2 &17.6 &100.0 &37.6 &26.4 &22.6 \\ \hline
 CIFAR100 &IN &98.0 &88.0 &64.0 &56.8 &98.0 &77.6 &47.2 &32.8 &98.0 &88.0 &68.0 &54.0 \\ \cline{3-14}
  &OUT &98.0 &73.4 &49.8 &37.8 &98.0 &68.4 &46.6 &34.4 &98.0 &80.0 &59.2 &42.2 \\ \hline
 REDDIT &EDGE &100.0 &39.3 &0.0 &0.0 &100.0 &0.0 &0.0 &0.0 &100.0 &16.0 &0.0 &0.0 \\ \hline
\end{tabular}
\vspace{-0.3cm}
\end{table*}


\noindent
{\bf Default configurations: } 
The default setup corresponds to a cross-device FL environment, where 10 clients are randomly selected per round. The total number of clients is 1,000 for the CIFAR10, DIGIT, and REDDIT, and 500 for CIFAR100 and BYCLASS. Client datasets are 
assigned using a Dirichlet distribution with an alpha of 0.9. The backdoor dataset comprises 512 examples, 60\% of which are correctly labeled and 40\% containing backdoor modifications. The only exception is with the FLShield defense, where the backdoor dataset is slightly smaller, containing approximately 300 legitimate examples and 50 to 80 malicious instances.

Unless stated otherwise, each experiment spans 1000 rounds of training. 
The first 300 rounds are under attack, while the subsequent 700  
operate with all clients behaving legitimately.

\subsection{MIGO under normal FL}
\label{sec_eval_MIGO}

This section examines MIGO under three scenarios. The Persistent adversary involves a single malicious client throughout the attack period. The Random adversary initially selects 1\% or 3\% of the participants to act maliciously, after which these clients may or may not be chosen by the server. This illustrates that the Random adversary has a considerably more constrained nature compared to the Persistent adversary.

Table 1 presents the \texttt{BackAcc} results for the different backdoor types. It displays the maximum backdoor accuracy observed during the experiment (MaxAcc). Additionally, it provides longevity values at various intervals: one hundred rounds after the attack concludes (L100), at the end of one attack duration (L300), and twice the attack period (L600).

\noindent
{\bf Backdoors were implanted:} The Persistent adversary successfully embedded all backdoors, with a very high \texttt{BackAcc} exceeding 90\%. This behavior remains largely consistent with Random adversaries, with only one instance where accuracy falls slightly below this threshold. The Random 1\% (3\%) adversary typically has around 30 (100) rounds where its clients are selected by the server. Consequently, between each backdoor injection attempt, there are, on average, several rounds where benign clients partially rectify the model. Nevertheless, the adversary can gradually accumulate modifications, leading the model to classify backdoor inputs as desired.

\begin{figure*}[!t]
\begin{minipage}[c]{0.3\linewidth}
  \centering
  \includegraphics[width=\columnwidth]{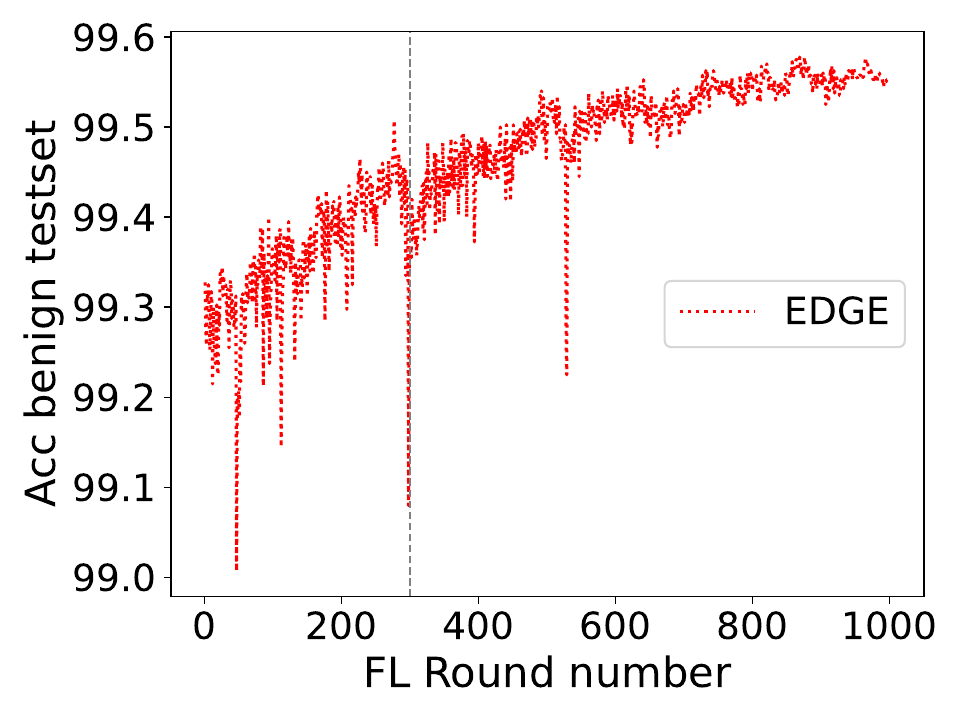}
  (a) Main task accuracy
\end{minipage}\hfill
\begin{minipage}[c]{0.3\linewidth}
  \centering
  \includegraphics[width=\columnwidth]{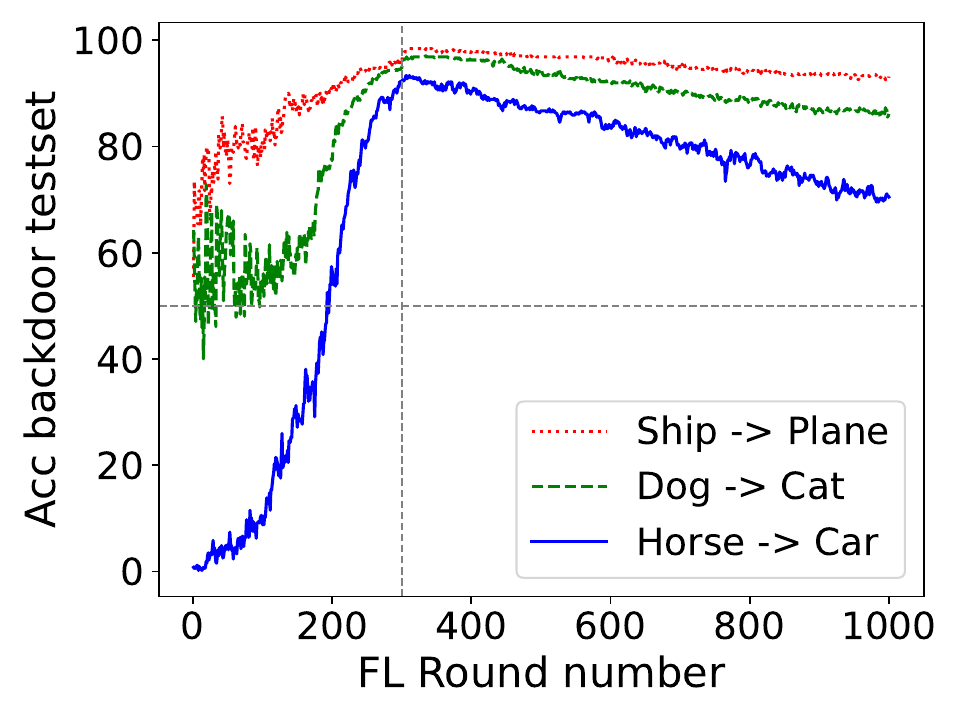}
  (b) OUT-Backdoors
\end{minipage}\hfill
\begin{minipage}[c]{0.3\linewidth}
  \centering
  \includegraphics[width=\columnwidth]{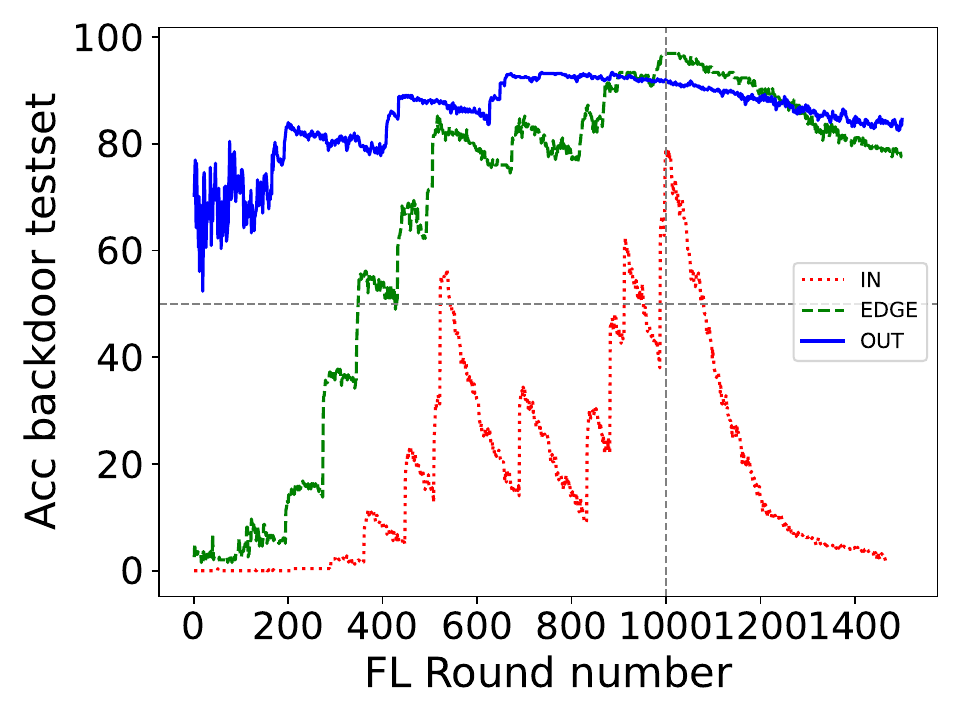}
  (c) Random 0.1\%
\end{minipage}\hfill
\caption{(a) Main task accuracy for EDGE backdoor while training with the DIGIT dataset;
(b) Three example OUT-backdoors with CIFAR10;
(c) An attack lasting 1000 rounds, with a Random adversary controlling 0.1\% of the participants (1 out 1000) with CIFAR10 and IN/EDGE/OUT backdoors;
 }
\label{fig_various_ND}
\end{figure*}

\noindent
{\bf Main task accuracy remains largely unaffected:} In the vast majority of datasets, the \texttt{BenAcc} remains largely unchanged during the attack interval, providing strong evidence of MIGO's stealthiness. This is exemplified in Figure~\ref{fig_various_ND}(a) with the EMNIST-DIGIT dataset, where benign accuracy steadily improves over 1000 rounds, increasing from 99.1\% to 99.5\%. The minor fluctuations observed are inherent to FL, as participants possess diverse datasets.
As previously explained, when IN backdoors are inserted, there is a slight decrease in \texttt{BenAcc} because the model begins to predict examples from the target class with the attacker-chosen label.

\begin{figure*}[!t]
\begin{minipage}[c]{0.3\linewidth}
  \centering
  \includegraphics[width=\columnwidth]{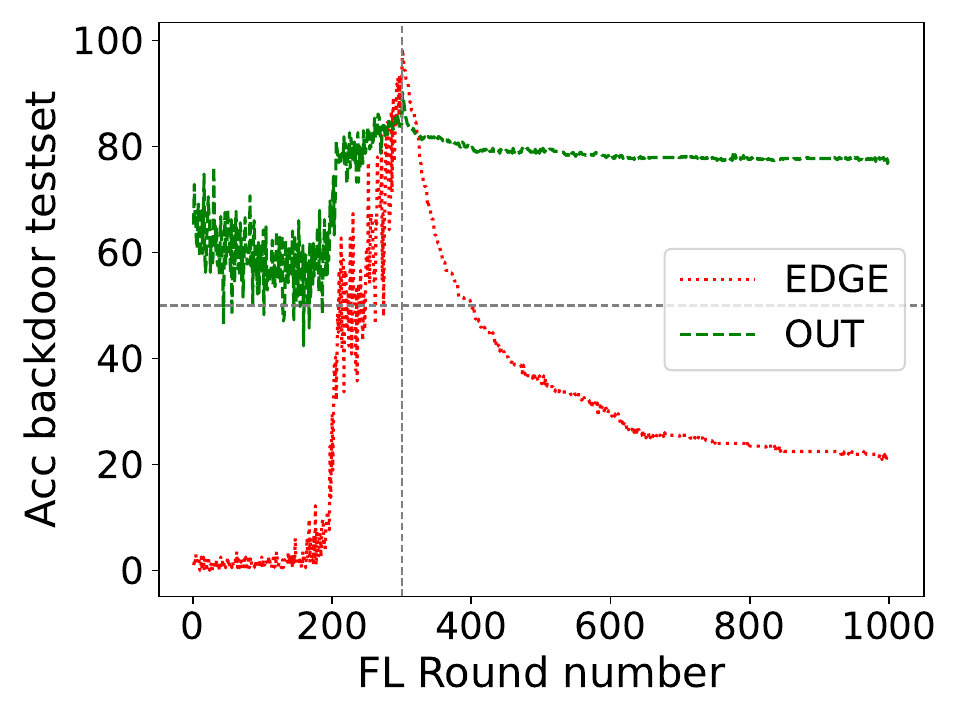}
  (a) Cross-silo with 20 clients
\end{minipage}\hfill
\begin{minipage}[c]{0.3\linewidth}
  \centering
  \includegraphics[width=\columnwidth]{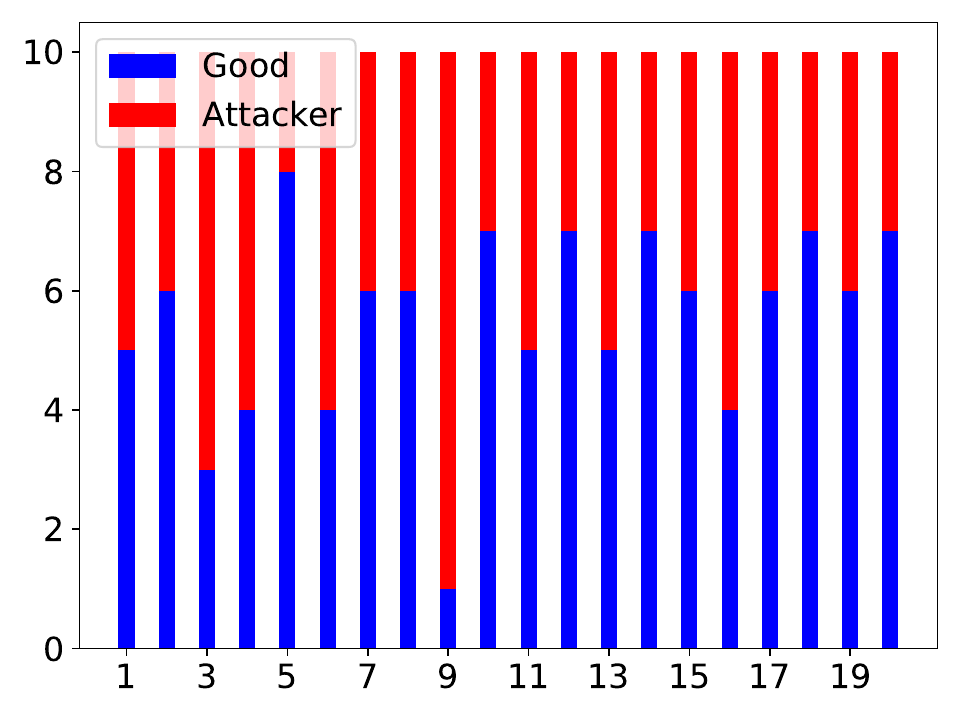}
  (b) Krum
\end{minipage}\hfill
\begin{minipage}[c]{0.3\linewidth}
  \centering
  \includegraphics[width=\columnwidth]{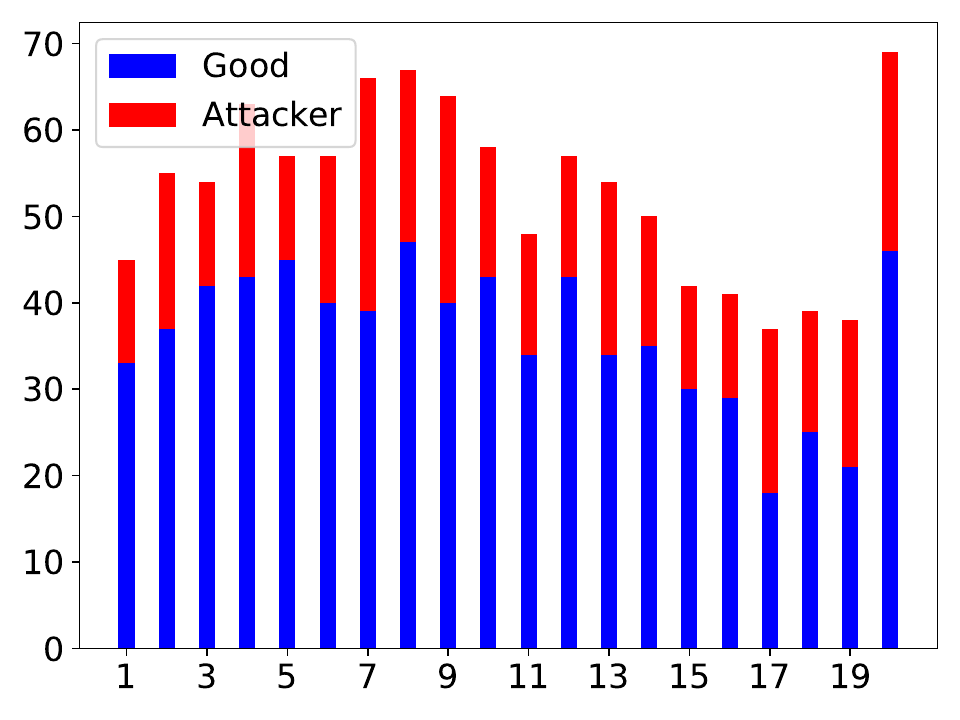}
  (c) FreqFed
\end{minipage}\hfill
\caption{(a) Cross-silo FL with 20 clients, where 4 are persistent attackers (EDGE/OUT backdoors with the CIFAR10 dataset).
(b) and (c) Number of malicious and benign client updates accepted by Krum and FreqFed over groups of 10 rounds (IN-backdoor during the first 200 rounds of the attack). }
\label{fig_defense}
\end{figure*}

\noindent
{\bf Longevity is significant in most scenarios:} With the Persistent adversary, longevity drops below 35\% in only two cases after 600 rounds (L600), while in half of the scenarios, it remains above 50\%. Conversely, results with the Random adversary are slightly lower due to fewer opportunities to embed the backdoors. As anticipated, adversaries controlling fewer nodes (Random setting) generally achieve worse outcomes.

In the CIFAR10 dataset, the EDGE and OUT backdoors outperform the IN backdoors due to fewer conflicts with updates from the correct nodes. However, in CIFAR100, the opposite trend is observed for OUT backdoors. This discrepancy is attributed to the challenge of training a model to predict 'camels' as 'clouds'. As shown in Figure~\ref{fig_various_ND}b, the selected OUT task can impact longevity (e.g., predicting 'ships' as 'planes' is less demanding than predicting 'horses' as 'cars'). Moreover, the task's difficulty influences the ease of backdoor insertion — in several cases, backdoor accuracy is very high (above 60\%) at the start of the attack. One possible explanation is that most features might have already been learned, allowing minimal changes to the model to enable a good prediction capability. These modifications also appear not to conflict with the main task, promoting backdoor persistence.

\noindent
{\bf Constrained adversaries can still be successful:} This experiment validates MIGO under challenging conditions, where a Random adversary only corrupts 0.1\% of the clients (i.e., 1 out of 1000). In this scenario, we extended the attack period to 1000 rounds, followed by an additional 500 rounds with only legitimate clients. Overall, we observed that the malicious client was only active around 15 times during the attack period.

The results are depicted in Figure~\ref{fig_various_ND} (c), showcasing the performance with the CIFAR10 dataset. Notably, whenever the malicious client executes, there is a small upward spike in \texttt{BackAcc}. Subsequently, during the intervals when only benign nodes participate in training, the accuracy slightly decreases, with a more pronounced decline observed for the IN backdoor. Ultimately, by the end of the attack, all three backdoors are successfully installed with high accuracies ranging from 79\% to 97\%. Longevity does not appear to be highly affected, particularly in the EDGE and OUT backdoors.

\noindent
{\bf Cross-silo attack scenarios are viable:} This experiment evaluates MIGO in a cross-silo scenario involving collaboration among 20 participants. Each client contributes to every round of training, but four participants behave maliciously during the initial 300 rounds. The datasets from each institution consist of 1024 samples, with approximately 100 examples per class of CIFAR10. The objective is to simulate a scenario where institutions possess reasonably sized datasets individually but are insufficient for standalone model training.

The results, illustrated in Figure~\ref{fig_defense} (a), depict a particularly challenging scenario where the attacker is significantly outnumbered in all attack rounds, and correct clients continuously provide updates to rectify the model across all predicted classes. Despite this, MIGO successfully embeds the EDGE/OUT backdoors with a max \texttt{BackAcc} above 90\%.


\begin{table*}[!t]
\renewcommand{\arraystretch}{1.2}
\setlength\tabcolsep{2pt}
\caption{Backdoor accuracy with MIGO and various defenses [CIFAR10 in cross-device configuration; 3 persistent attackers].}
\small
\label{tab_defense}
\centering
\begin{tabular}{|c|cccc|cccc|cccc|} \cline{2-13}
\multicolumn{1}{c|}{}  & \multicolumn{4}{c|}{IN} & \multicolumn{4}{c|}{EDGE} & \multicolumn{4}{c|}{OUT} \\ \cline{1-1}
 Defense     &MaxAcc    &L100      &L300      &L600      &MaxAcc    &L100      & L300      &L600     &MaxAcc     &L100    &L300   &L600 \\ \hline \hline
 Norm Clipping (NC) &96.6 &92.2 &79.6 &54.0 &100.0 &100.0 &98.0 &95.9 &97.7 &97.1 &93.2 &86.8 \\ \cline{2-13}
 NC+Noise &97.0 &93.6 &81.2 &55.0 &100.0 &100.0 &98.4 &94.1 &97.9 &97.2 &93.4 &89.4 \\ \cline{2-13}
 Krum &94.2 &82.6 &52.5 &16.7 &100.0 &99.3 &97.4 &95.3 &96.7 &93.2 &91.7 &87.2 \\ \cline{2-13}
 mKrum &80.6 &62.4 &25.1 &8.2 &98.0 &98.0 &98.0 &94.4 &91.4 &89.6 &87.9 &85.1 \\ \cline{2-13}
 Flame &88.6 &79.8 &41.7 &13.1 &99.0 &98.0 &96.9 &94.6 &90.8 &89.8 &87.0 &83.4 \\ \cline{2-13}
 FoolsGold &90.4 &88.8 &81.4 &59.8 &100.0 &99.5 &98.9 &98.0 &96.5 &93.4 &82.2 &81.9 \\ \cline{2-13}
 DeepSight &71.6 &45.8 &23.4 &7.4 &98.5 &98.5 &96.4 &93.6 &90.8 &90.2 &87.2 &82.8 \\ \cline{2-13}
 CrowdGuard &96.0 &85.6 &60.8 &27.9 &100.0 &98.0 &94.8 &84.7 &97.3 &92.5 &84.1 &80.4 \\ \cline{2-13}
 FreqFed &92.2 &86.6 &60.0 &14.4 &99.0 &90.7 &83.0 &73.4 &96.5 &84.4 &81.6 &77.2 \\ \cline{2-13}
 FLShield &81.8 &60.9 &16.6 &0.0 &98.5 &96.9 &93.3 &84.1 &93.2 &89.7 &86.6 &83.4 \\ \hline
\end{tabular}
\end{table*}


\subsection{MIGO with Defenses}
\label{sec_defenses}

This section evaluates MIGO against a diverse array of defenses, including several state-of-the-art methods known for their high effectiveness in mitigating various poison attacks. Furthermore, it examines MIGO's ability to bypass these defenses by leveraging its tailored evasion mechanisms. Most of the defenses considered assume that the attacker controls less than half of the participants ($n_a + n_b = n; n_a < n/2$). However, Krum and mKrum~\cite{krum} impose a stricter requirement ($2n_a + 2 < n$), which we adhered to in all experiments for fairness. Consequently, the evaluation was conducted using 3 persistent attackers out of the 10 round participants, with the CIFAR10 dataset.

The results are summarized in Table~\ref{tab_defense}. In all cases, it is possible to install the three types of backdoors, in most cases with a very high level of accuracy. The EDGE and OUT backdoors show strong longevity, maintaining significant accuracy over 600 rounds after the attack ceased. In contrast, IN backdoors proved more challenging to install and sustain, as benign clients systematically propose corrections to the global model during training because their datasets include examples of the class under attack. 

Experiments show that MIGO could continuously mislead the defenses and gradually implant backdoors over the 300 rounds of the attack. While defenses sometimes exclude or penalize a subset (or even all) of the malicious updates, this typically occurs only for a limited number of rounds. In addition, defenses also frequently reject legitimate updates, thereby restricting valid changes to the global model and facilitating the attack. For instance, Figure~\ref{fig_defense}c illustrates the number of benign and malicious updates accepted by FreqFed\cite{freqfed} during the first 200 rounds of the attack, with each bar representing accumulated values over 10 rounds. The average number of benign updates is significantly lower than the expected 70 with perfect detection, and the malicious attackers consistently evade FreqFed by a relevant margin.

Considering the specific mechanisms in more detail:

\noindent
{\bf Backdoor types and mixed training dataset.} 
DeepSight~\cite{deepsight} makes participant models process a random dataset and computes a specialized metric (DDifs) to cluster model updates, ensuring that all models within a group have been trained on similar datasets. Conversely, CrowdGuard~\cite{crowdguard} engages round participants to evaluate the submitted models with their datasets to collect the outputs of the layers. These outputs are then compared (using cosine and Euclidean distances) with those of the global model, and the results are examined through several statistical tests to identify poisoned models. 

These analyses are imprecise when there is variability among datasets, which causes local models to become much less consistent. Moreover, the backdoor samples lack special triggers and share features with normal input instances. Attackers also use a training dataset that contains a majority of valid and correctly labeled samples. These attributes make it more challenging for defenses to distinguish malicious local models from legitimate ones, thus contributing to attack stealthiness.

\noindent
{\bf Projecting the model around the good.} ESR constrains the exploration of the parameter space to a roughly defined area around the current global model, $\mathcal{G}_{r}$. If set too small, usually it is unfeasible to find a path to minima with high longevity. Conversely, if set too large, SGD may move the attacker model too far, causing a decrease in benign accuracy after aggregation. Therefore, ESR was kept constant to an intermediate value of 3.0. This value produced positive results across all defenses, indicating a good level of robustness. 

MPR ensures that the submitted attacker models remain close to the global model. However, the extent of this region should be tailored to the specific defense to maximize attack impact. In some cases, it is straightforward to select the optimal value — e.g., with NC or NC+Noise~\cite{SunBackdoor19}, the MPR value should match the threshold applied by the defense (0.2 in these experiments). 

Other defenses do not look for distances in the parameter space, and therefore, they are less sensitive to the MPR. In those cases, we selected a bound of 0.3 and kept it static during the whole experiment. This applies, for instance, to CrowdGuard~\cite{crowdguard}, which focuses the analysis on the layer outputs of local models.

Defenses like Krum~\cite{krum}, Flame~\cite{flame}, and FLShield~\cite{flshield} use metrics (L2 norm; cosine distance) to measure the gap between local models. They may also examine the separation between the local models and the current global model. When facing such defenses, the adversary needs to continuously adjust the MPR to ensure that the attacker's models blend with the legitimate ones. This is challenging for attacks that last for many rounds, as distances tend to diminish. MIGO uses an estimation procedure (Algorithm~\ref{alg_migo}) to address the issue, with a defense-specific $\beta$ factor. For example, Krum selects only a single model per round for aggregation --- the model closest to $n-f-2$ other models (where $f$ is the presumed number of malicious clients). Consequently, successful attacks require that poisoned models are frequently located near the "center". This could be achieved with a static $\beta$ of 1.6 (see Figure~\ref{fig_defense}b). In one case, MIGO used a dynamic $\beta$ factor to limit the decrease of MPR as the attack progressed. This occurred with FLShield and the IN backdoor. 

Lastly, some defenses apply a transformation to the parameter space before calculating distances to highlight specific patterns unique to malicious models. FreqFed~\cite{freqfed} is a prime example of this approach. It employs a Discrete Cosine Transform to enable analysis in the frequency domain, where it looks for energy shifts in the low-frequency components of the spectrum. MIGO, however, still mixed malicious and benign models, repeatedly ensuring that a subset of the attacker updates were aggregated (see Figure~\ref{fig_defense}c).

\noindent
{\bf Layer forcing.} The method enables specific layers of the attacker models to mimic those found in updates by correct clients. This is achieved by constraining, during training, selected layers to resemble those in a legitimate model. Furthermore, the adversary can calibrate the similarity among attacker models by adjusting the learning of the benign models being impersonated (e.g., running SGD through more batches).

FoolsGold~\cite{foolsgold} searches for Sybil clones by comparing the cosine similarity between the output layer weights of the submitted models and their historical values, to measure the level of disparity among pairs of models. The adversary can counter this by extending the training of the benign models for a few extra epochs to increase the divergence among the output layers. This approach contributes to enhancing the likelihood of malicious updates being accepted. 

DeepSight~\cite{deepsight} evaluates the magnitudes of updates for individual neurons in the output layer to estimate the label distribution in the training data. This approach helps to identify models that concentrate most of the learning in a single (attack) class.
In this case, the adversary conducts benign model training similar to that of normal clients. Overall, this technique successfully generated malicious models that were frequently perceived as legitimate.


\begin{table*}[!t]
\renewcommand{\arraystretch}{1.2}
\setlength\tabcolsep{2pt}
\caption{Backdoor accuracy for five attack strategies and three defense approaches [CIFAR10; 3 persistent attackers].}
\small
\label{tab_attack}
\centering
\begin{tabular}{|cc|cccc|cccc|cccc|} \cline{3-14}
\multicolumn{2}{c|}{}  & \multicolumn{4}{c|}{Norm Clipping} & \multicolumn{4}{c|}{FreqFed} & \multicolumn{4}{c|}{FLShield} \\ \cline{1-2}
 Backd   &Attack     &MaxAcc    &L100      &L300      &L600      &MaxAcc    &L100      & L300      &L600     &MaxAcc     &L100    &L300   &L600 \\ \hline \hline
  &BackPGD &93.6 &17.9 &4.6 &0.0 &95.4 &62.6 &45.3 &29.8 &3.2 &0.0 &0.0 &0.0 \\ \cline{3-14}
  &MRepl &91.8 &0.0 &0.0 &0.0 &89.4 &65.5 &38.2 &15.8 &0.6 &0.0 &0.0 &0.0 \\ \cline{3-14}
 IN &Neuro &97.6 &92.8 &73.5 &12.8 &0.2 &0.0 &0.0 &0.0 &1.0 &0.0 &0.0 &0.0 \\ \cline{3-14}
  &3DFed &55.6 &15.9 &2.7 &0.0 &0.2 &0.0 &0.0 &0.0 &80.2 &37.7 &9.5 &2.3 \\ \cline{3-14}
  &MIGO &97.2 &92.3 &77.4 &47.0 &90.4 &87.7 &57.5 &13.3 &81.8 &60.9 &16.6 &0.0 \\ \hline \hline
  &BackPGD &99.5 &81.6 &54.8 &45.1 &99.5 &46.2 &40.2 &32.1 &100.0 &90.8 &80.3 &63.4 \\ \cline{3-14}
  &MRepl &100.0 &41.5 &34.1 &30.5 &6.1 &2.8 &3.1 &2.6 &4.6 &3.1 &3.1 &3.1 \\ \cline{3-14}
 EDGE &Neuro &100.0 &98.5 &93.3 &75.5 &100.0 &97.4 &95.9 &88.1 &96.9 &58.6 &58.3 &52.2 \\ \cline{3-14}
  &3DFed &89.3 &80.2 &67.9 &55.4 &4.1 &2.0 &2.0 &2.1 &98.5 &90.8 &80.5 &64.1 \\ \cline{3-14}
  &MIGO &100.0 &100.0 &98.0 &94.3 &99.5 &97.4 &92.9 &86.0 &98.5 &96.9 &93.3 &84.1 \\ \hline \hline
  &BackPGD &95.1 &80.8 &82.9 &83.8 &96.1 &87.5 &87.0 &81.5 &98.6 &86.4 &87.2 &85.8 \\ \cline{3-14}
  &MRepl &98.2 &75.9 &77.5 &77.8 &94.3 &79.4 &71.1 &64.8 &87.1 &68.5 &70.1 &68.6 \\ \cline{3-14}
 OUT &Neuro &98.8 &95.5 &83.2 &77.5 &80.5 &64.9 &66.5 &65.4 &83.0 &67.1 &67.9 &71.1 \\ \cline{3-14}
  &3DFed &88.7 &85.8 &80.2 &74.5 &76.0 &66.1 &66.6 &66.2 &93.8 &91.4 &87.2 &81.4 \\ \cline{3-14}
  &MIGO &99.2 &96.4 &94.1 &86.2 &96.7 &83.2 &82.3 &81.4 &93.2 &89.7 &86.6 &83.4 \\ \hline
\end{tabular}

\end{table*}


\vspace{-0.3cm}
\subsection{MIGO vs. Other attack strategies} \label{sec_attacks}

This section evaluates the performance of MIGO against state-of-the-art backdoor insertion strategies~\cite{SunBackdoor19,mrepl,neurotoxin,3dfed} in circumventing three defense approaches. These defenses were chosen to provide a comprehensive assessment: one serves as a baseline~\cite{SunBackdoor19}, while the others are recent, highly effective techniques~\cite{freqfed,flshield}. Each defense represents a different method for mitigating corrupted model updates, facilitating a robust comparison of the attack strategies. We performed a hyperparameter search to optimize each attack, using the configurations from the original studies as starting points. Table~\ref{tab_attack} summarizes the results, revealing two main insights:

First, MIGO is the only approach that successfully evades all defenses, maintaining high accuracy while embedding backdoors with lasting impact. More advanced defenses, such as FreqFed and FLShield, typically discard only a subset of the local models sent by attacker clients, allowing the remaining ones to systematically contaminate the global model.

Second, the OUT backdoor poses a notable risk to FL systems due to its high level of success across the attack and defense strategies. As discussed in Section~\ref{sec_eval_MIGO}, this occurs because OUT backdoor examples ({\it dogs}) are assigned by the model to the adversary's target class ({\it cats}) with considerable probability. This finding suggests that depending on the adversary's goals, it may be intrinsically challenging to prevent such behavior, potentially underscoring the need for alternative defense mechanisms in the future.

Detailed findings on each of the defenses are as follows:

\noindent {\bf Norm clipping defense (NC):} NC demonstrated limited effectiveness in preventing global model abuse, given the strict threshold of 0.4 that was used. As NC does not filter updates but rather constrains their influence on the global model, the attacks could continuously implant backdoors during attack rounds, achieving strong accuracy levels. However, NC was capable of limiting lasting changes with the IN backdoor and three strategies ---BackPGD~\cite{SunBackdoor19}, MRepl~\cite{mrepl}, and 3DFed~\cite{3dfed}--- as benign clients quickly fixed the global parameters after the attack ceased.

\noindent {\bf FreqFed defense:} Assuming the adversary can find the region where benign updates occur, BackPGD can set up attacker clients to train with projected gradient descent within this same general area. Consequently, BackPGD could successfully mislead FreqFed throughout most of the initial 180 attack rounds. Notably, in some rounds, only malicious updates were selected by FreqFed, as benign updates either remained unassigned to clusters or were placed to minority clusters --- an effect stemming from the use of HDBSan~\cite{HDBSan}. This enabled BackPGD to successfully implant all backdoors.

MRepl showcases the classic dilemma, where \emph{"defenders have to be right 100\% of the time, whereas attackers only need to succeed once"}. Over the 300 attack rounds, FreqFed successfully eliminated every malicious update in all but one round. However, in that crucial round, FreqFed mistakenly accepted all three bad updates while excluding every benign model. Since attacker clients applied a boosting factor of 3 to their updates, this enabled a near-complete model replacement. This enabled the embedding of the IN and OUT backdoors but was less effective for EDGE. 
Neuro~\cite{neurotoxin} experienced a similar fortuitous event with the EDGE backdoor, which was enough to successfully poison the global model. For the other types of backdoors, all malicious updates were filtered out. 

3DFed was unable to evade FreqFed. Upon detecting that its malicious updates were consistently rejected, 3DFed attempted to adapt its injection strategy. When these alternatives also proved ineffective, it disabled the adaptive tuning approach and continued unsuccessfully trying to backdoor the global model until the attack concluded. This outcome was anticipated, as 3DFed lacks specific evasion mechanisms to counter FreqFed’s frequency-domain analysis.

\noindent {\bf FLShield defense:} Similar to FreqFed, BackPGD attempted to project corrupted updates in the same region as those of benign clients. However, this was ineffective with the IN backdoor. Since the IN backdoor mislabels one class, it results in a significant loss difference between the cluster's representative model and the previous round global model on examples of the backdoored class. The LIPC metric detects this discrepancy, leading to the removal of all updates in the cluster containing the malicious clients. With EDGE and OUT backdoors, FLShield also managed to cluster attacker updates together. Yet, in over half of the attack rounds, the cluster associated with attackers was allowed to update the global model due to its relatively low LIPC.

MRepl was ineffective at installing any backdoors, as the attacker updates were grouped into a single cluster and subsequently excluded due to their high LIPC scores. As FLShield sometimes eliminates many model updates per round, there were occasional spikes in \texttt{BackAcc} during training with the OUT backdoor, raising \texttt{MaxAcc} to around 87\%.

Neuro exhibited a similar pattern to BackPGD, ultimately proving ineffective with the IN backdoor. In the EDGE scenario, there were only a few tens of rounds where attacker models bypassed filtering, allowing for backdoor installation, though with somewhat limited persistence. For the OUT backdoor, malicious updates were rarely selected, as they generally had higher LIPC values compared to benign updates.

3DFed could, to a certain degree, evade FLShield. 3DFed includes a constraint in its loss function that promotes cosine similarity and minimizes the Euclidean distance between the attacker’s models and the global model. It also adds noise masks to the attacker models, enhancing their dissimilarity. Since FLShield clusters models based on cosine distances, these steps caused attacker models to be distributed in different clusters. Additionally, the LIPC metric for these clusters was not significant. As a result, in most rounds, filtering excluded only a subset of poisoned models, allowing the remaining ones to inject the backdoor.

\section{Related Work}
\label{related_work}

This section reviews the literature on attacks and defenses in federated learning, with an emphasis on backdoor attacks.

\paragraph{Attacks.} 

Several attack methods have been developed for FL, and sometimes they may be combined to maximize impact. Some of these attacks can target the overall inference capability of the global model when training concludes (\textit{untargeted poison attacks}) or change the model's behavior for specific classes (\textit{targeted poison attacks}).

\emph{- Model/Data poisoning:} With model-poisoning, the adversary manipulates the client-side training process by altering hyperparameters, modifying the loss function~\cite{mrepl, Bhagoji19}, or directly changing model updates~\cite{Baruch19, Fang20, Shejwalkar21}. With data poisoning, the attacker changes the training data on a subset of clients, altering datasets to achieve specific goals, such as reducing accuracy for particular classes~\cite{Tolpegin20} or forcing the model to output a chosen label when a trigger is present in the input~\cite{Xie19}. Data poisoning may be easier to execute than model poisoning, as it only requires compromising the device’s storage rather than the FL procedure itself. Moreover, detection methods based on example inspection are challenging to implement in FL, as clients are not required to share their data.

\emph{- Backdoor attacks:} Backdoor attacks aim to alter the global model’s behavior for specific inputs or classes~\cite{SunBackdoor19,mrepl,Wang20,Xie19,Bhagoji19,neurotoxin,3dfed}, making them targeted attacks. They can be introduced into the global model through various strategies. 
For example, a global backdoor trigger can be split into distinct patterns, each embedded in the local datasets of different malicious participants~\cite{Xie19}. This approach ensures that the global model will respond as intended by the adversary when it encounters the complete trigger pattern. Alternatively, attackers can inject edge-case backdoors~\cite{Wang20,mrepl}, causing the model to misclassify examples on the tail end of the input distribution. In a model replacement attack, the adversary attempts to substitute the global model entirely with a malicious local update (as in MRepl~\cite{mrepl}), exploiting the assumption that, as the global model converges, local (benign) updates start to cancel out. This cancellation creates an opening for a boosted (malicious) update to take over the global model. Boosting can also be combined with other adversarial goals, such as stealth, by framing it as an optimization problem solved during local training~\cite{Bhagoji19,Kraub24}. Furthermore, local training can be tuned to ensure updates remain close to the global model (as in BackProj~\cite{SunBackdoor19}) or to ensure gradients align within the bottom-K\% of benign coordinates (as in Neurotoxin~\cite{neurotoxin}). The attack can also be made adaptive, employing various countermeasures to evade particular defenses in a black-box setting (as in 3DFed~\cite{3dfed}).

MIGO aims to generate malicious updates that introduce sufficient ambiguity, making it challenging for defenses to reliably distinguish between legitimate and malicious local models. This strategy is implemented through several techniques, such as choosing backdoor types that closely resemble normal examples and incrementally introducing small updates to the global model.

\paragraph{Protections.} Defenses in machine learning models typically focus on filtering attacks and/or mitigating their impact. Although the literature has attempted to categorize defenses, achieving this goal is challenging due to the fact that state-of-the-art safeguards often integrate multiple approaches. 

\emph{- Robust aggregation}: FedAvg uses the mean as the statistical operator to aggregate client updates, making it susceptible to outliers. One way to address this limitation is to use other operators, such as median or geometric-median~\cite{Pillutla_defesa20}. Trimmed-mean is another alternative, which filters extreme values below and above the data distribution by a certain percentage and then calculates the mean of the remaining values~\cite{Yin_defesa18}. Other robust aggregation solutions have been proposed, for example, CRFL~\cite{Xie21}. 

\emph{- Clipping}: Norm clipping thwarts attacks that increase the norm of the updates to amplify impact~\cite{SunBackdoor19}. NC disregards client updates above a predefined threshold value or constrains the norm of the update to that value. However, there are limitations --- a determined attacker might adjust updates and evade the defense, and a fixed threshold may be bypassed through distributed attacks~\cite{Xie19}. To address these concerns, Guo et al. proposed a dynamic clipping approach that adjusts the threshold during training~\cite{Guo21}. Additionally, clipping techniques have been proven effective in various practical scenarios, as demonstrated by Shejwalkar et al.~\cite{Shejwalkar22}.

\emph{- Detection and filtering}: Generically, this defense analyzes the updates and eliminates the suspicious before the aggregation~\cite{FLDetector,Andreina_defesa21,BayBFed}. For example, Krum measures the Euclidean distance from each client update to the \emph{n-f-2} nearest neighbors and then removes the updates with the highest distances~\cite{krum}. To decrease the impact of Sybil-based poisoning, FoolsGold finds the gradient updates most similar to each other (using cosine similarity) and decreases the weights associated with their contributions~\cite{foolsgold}. Other approaches evolved from these ideas to identify (and discard) malicious updates by combining several metrics together with statistical tests~\cite{MESAS,crowdguard}. Hybrid solutions have also been presented in the literature, employing a combination of approaches to eliminate potential malicious updates or to reduce their influence in the global model~\cite{FLTrust,deepsight,flame,flshield}. Filtering methods, however, encounter a fundamental challenge in FL to accurately distinguish between two updates that differ due to one being malicious or because they originate from clients with very diverse datasets. This problem leads to filtering errors that undermine benign accuracy, and adaptive attackers can explore them to escape detection~\cite{Shejwalkar21}.

\emph{- Noise/Differential Privacy}: These solutions typically involve adding random Gaussian noise to model updates~\cite{Brendan18,deep_learning_diff_privacy}, thereby reducing the influence of specific malicious data points. This mechanism has been applied at the server to prevent the injection of backdoors~\cite{Xie21,flame,SunBackdoor19}, as well as at clients to counter attacks aimed at inferring information from observed model updates~\cite{Wei20}.



\section{Conclusions}
Our research presents MIGO, an innovative method for embedding backdoors into machine learning models trained in FL. MIGO strategically selects backdoor types with inputs that naturally blend with legitimate examples. Backdoors are integrated progressively throughout the training process, enabling the generation of malicious model updates that closely resemble benign ones. MIGO successfully implanted distinct types of backdoors, even in the presence of robust and diverse FL defenses. These results highlight the substantial threat posed by this attack, which achieves high backdoor accuracy while preserving the utility of the main task.

\section*{Acknowledgments}

This work was supported by FCT through the LASIGE Research Unit, ref. UID/000408/2025.


\bibliographystyle{IEEEtran}
\bibliography{FL_biblio.bib}

\begin{thebibliography}{10}
\providecommand{\url}[1]{#1}
\csname url@samestyle\endcsname
\providecommand{\newblock}{\relax}
\providecommand{\bibinfo}[2]{#2}
\providecommand{\BIBentrySTDinterwordspacing}{\spaceskip=0pt\relax}
\providecommand{\BIBentryALTinterwordstretchfactor}{4}
\providecommand{\BIBentryALTinterwordspacing}{\spaceskip=\fontdimen2\font plus
\BIBentryALTinterwordstretchfactor\fontdimen3\font minus \fontdimen4\font\relax}
\providecommand{\BIBforeignlanguage}[2]{{%
\expandafter\ifx\csname l@#1\endcsname\relax
\typeout{** WARNING: IEEEtran.bst: No hyphenation pattern has been}%
\typeout{** loaded for the language `#1'. Using the pattern for}%
\typeout{** the default language instead.}%
\else
\language=\csname l@#1\endcsname
\fi
#2}}
\providecommand{\BIBdecl}{\relax}
\BIBdecl

\bibitem{FedAvg}
H.~McMahan, E.~Moore, D.~Ramage, S.~Hampson, and B.~Arcas, ``Communication-efficient learning of deep networks from decentralized data,'' in \emph{Proceedings of the the International Conference on Artificial Intelligence and Statistics}, 2017, pp. 1273--1282.

\bibitem{fl_design}
K.~Bonawitz \emph{et~al.}, ``Towards federated learning at scale: System design,'' in \emph{Proceedings of the Machine Learning and Systems}, 2019, pp. 374--388.

\bibitem{GDPR2016a}
\BIBentryALTinterwordspacing
{European Parliament} and {Council of the European Union}, ``Regulation ({EU}) 2016/679 of the {European} {Parliament} and of the {Council} of 27 {April} 2016 on the protection of natural persons with regard to the processing of personal data and on the free movement of such data, and repealing {Directive} 95/46/{EC} ({General} {Data} {Protection} {Regulation}).'' [Online]. Available: \url{https://data.europa.eu/eli/reg/2016/679/oj}
\BIBentrySTDinterwordspacing

\bibitem{Bukaty19}
\BIBentryALTinterwordspacing
P.~Bukaty, \emph{The California Consumer Privacy Act (CCPA): An implementation guide}.\hskip 1em plus 0.5em minus 0.4em\relax IT Governance Publishing, 2019. [Online]. Available: \url{http://www.jstor.org/stable/j.ctvjghvnn}
\BIBentrySTDinterwordspacing

\bibitem{autonomous_driving_fl}
A.~M. Elbir, B.~Soner, S.~{\c{C}}{\"o}leri, D.~G{\"u}nd{\"u}z, and M.~Bennis, ``Federated learning in vehicular networks,'' in \emph{Proceedings of the IEEE International Mediterranean Conference on Communications and Networking}, 2022, pp. 72--77.

\bibitem{healthcare_fl}
M.~Joshi, A.~Pal, and M.~Sankarasubbu, ``Federated learning for healthcare domain - pipeline, applications and challenges,'' \emph{ACM Transactions on Computing for Healthcare}, vol.~3, no.~4, 2022.

\bibitem{gboard_fl}
\BIBentryALTinterwordspacing
T.~Yang, G.~Andrew, H.~Eichner, H.~Sun, W.~Li, N.~Kong, D.~Ramage, and F.~Beaufays, ``Applied federated learning: Improving google keyboard query suggestions,'' \emph{CoRR}, vol. abs/1812.02903, 2018. [Online]. Available: \url{http://arxiv.org/abs/1812.02903}
\BIBentrySTDinterwordspacing

\bibitem{siri_fl}
\BIBentryALTinterwordspacing
K.~Hao, ``{How Apple personalizes Siri without hoovering up your data},'' December 2019, [Online; accessed 8-November-2024]. [Online]. Available: \url{\url{https://www.technologyreview.com/2019/12/11/131629/apple-ai-personalizes-siri-federated-learning/ }}
\BIBentrySTDinterwordspacing

\bibitem{Tolpegin20}
V.~Tolpegin, S.~Truex, M.~Gursoy, and L.~Liu, ``Data poisoning attacks against federated learning systems,'' in \emph{Proceedings of the European Symposium on Research in Computer Security}, 2020, pp. 480--501.

\bibitem{Shejwalkar21}
V.~Shejwalkar and A.~Houmansadr, ``Manipulating the byzantine: Optimizing model poisoning attacks and defenses for federated learning,'' in \emph{Proceedings of the Network and Distributed System Security Symposium}, 2021, pp. 18--37.

\bibitem{Fang20}
M.~Fang, X.~Cao, J.~Jia, and N.~Gong, ``Local model poisoning attacks to byzantine-robust federated learning,'' in \emph{Proceedings of the USENIX Conference on Security Symposium}, 2020, pp. 1623--1640.

\bibitem{neurotoxin}
Z.~Zhang, A.~Panda, L.~Song, Y.~Yang, M.~Mahoney, J.~Gonzalez, K.~Ramchandran, and P.~Mittal, ``Neurotoxin: Durable backdoors in federated learning,'' in \emph{Proceedings of the International Conference on Machine Learning}, 2022, pp. 26\,429--26\,446.

\bibitem{Bhagoji19}
A.~Bhagoji, S.~Chakraborty, P.~Mittal, and S.~Calo, ``Analyzing federated learning through an adversarial lens,'' in \emph{Proceedings of the International Conference on Machine Learning}, Jun 2019, pp. 634--643.

\bibitem{mrepl}
E.~Bagdasaryan, A.~Veit, Y.~Hua, D.~Estrin, and V.~Shmatikov, ``How to backdoor federated learning,'' in \emph{Proceedings of the International Conference on Artificial Intelligence and Statistics}, 2020, pp. 2938--2948.

\bibitem{3dfed}
H.~Li, Q.~Ye, H.~Hu, J.~Li, L.~Wang, C.~Fang, and J.~Shi, ``3dfed: Adaptive and extensible framework for covert backdoor attack in federated learning,'' in \emph{Proceedings of the IEEE Symposium on Security and Privacy}, 2023, pp. 1893--1907.

\bibitem{SunBackdoor19}
Z.~Sun, P.~Kairouz, A.~T. Suresh, and H.~B. McMahan, ``Can you really backdoor federated learning?'' \emph{CoRR}, vol. abs/1911.07963, 2019.

\bibitem{Wang20}
H.~Wang, K.~Sreenivasan, S.~Rajput, H.~Vishwakarma, S.~Agarwal, J.-Y. Sohn, K.~Lee, and D.~Papailiopoulos, ``Attack of the tails: Yes, you really can backdoor federated learning,'' in \emph{Proceedings of the International Conference on Neural Information Processing Systems}, 2020, pp. 16\,070--16\,084.

\bibitem{Xie19}
C.~Xie, K.~Huang, P.-Y. Chen, and B.~Li, ``Dba: Distributed backdoor attacks against federated learning,'' in \emph{Proceedings of the International Conference on Learning Representations}, 2019.

\bibitem{fldefinitions}
P.~Kairouz \emph{et~al.}, ``Advances and open problems in federated learning,'' \emph{Foundations and Trends in Machine Learning}, vol.~14, no. 1-2, pp. 1--210, 2021.

\bibitem{Gu17}
\BIBentryALTinterwordspacing
T.~Gu, B.~Dolan{-}Gavitt, and S.~Garg, ``{BadNets}: Identifying vulnerabilities in the machine learning model supply chain,'' \emph{CoRR}, vol. abs/1708.06733, 2017. [Online]. Available: \url{http://arxiv.org/abs/1708.06733}
\BIBentrySTDinterwordspacing

\bibitem{Chen17}
X.~Chen, C.~Liu, B.~Li, K.~Lu, and D.~Song, ``Targeted backdoor attacks on deep learning systems using data poisoning,'' \emph{CoRR}, vol. abs/1712.05526, 2017.

\bibitem{Suciu18}
O.~Suciu, R.~Marginean, Y.~Kaya, H.~D. III, and T.~Dumitras, ``When does machine learning {FAIL}? generalized transferability for evasion and poisoning attacks,'' in \emph{Proceedings of the USENIX Security Symposium}, Aug. 2018, pp. 1299--1316.

\bibitem{Turner2019}
A.~Turner, D.~Tsipras, and A.~Madry, ``Label-consistent backdoor attacks,'' \emph{CoRR}, vol. abs/1912.02771, 2019.

\bibitem{Saha2020}
A.~Saha, A.~Subramanya, and H.~Pirsiavash, ``Hidden trigger backdoor attacks,'' \emph{Proceedings of the AAAI Conference on Artificial Intelligence}, vol.~34, no.~07, pp. 11\,957--11\,965, Apr. 2020.

\bibitem{Bagdasaryan21}
E.~Bagdasaryan and V.~Shmatikov, ``Blind backdoors in deep learning models,'' in \emph{Proceedings of the USENIX Security Symposium}, Aug. 2021, pp. 1505--1521.

\bibitem{Chen21}
X.~Chen, A.~Salem, D.~Chen, M.~Backes, S.~Ma, Q.~Shen, Z.~Wu, and Y.~Zhang, ``{BadNL}: Backdoor attacks against {NLP} models with semantic-preserving improvements,'' in \emph{Proceedings of the Annual Computer Security Applications Conference}, 2021, pp. 554--569.

\bibitem{Pan22}
X.~Pan, M.~Zhang, B.~Sheng, J.~Zhu, and M.~Yang, ``Hidden trigger backdoor attack on {NLP} models via linguistic style manipulation,'' in \emph{Proceedings of the USENIX Security Symposium}, Aug. 2022, pp. 3611--3628.

\bibitem{Boucher22}
N.~Boucher, I.~Shumailov, R.~Anderson, and N.~Papernot, ``Bad characters: Imperceptible {NLP} attacks,'' in \emph{Proceedings of the IEEE Symposium on Security and Privacy}, 2022, pp. 1987--2004.

\bibitem{Gao20}
Y.~Gao, B.~G. Doan, Z.~Zhang, S.~Ma, J.~Zhang, A.~Fu, S.~Nepal, and H.~Kim, ``Backdoor attacks and countermeasures on deep learning: {A} comprehensive review,'' \emph{CoRR}, vol. abs/2007.10760, 2020.

\bibitem{Li24}
Y.~Li, Y.~Jiang, Z.~Li, and S.-T. Xia, ``Backdoor learning: A survey,'' \emph{IEEE Transactions on Neural Networks and Learning Systems}, vol.~35, no.~1, pp. 5--22, 2024.

\bibitem{flshield}
E.~Kabir, Z.~Song, M.~R. Ur~Rashid, and S.~Mehnaz, ``{ FLShield: A Validation Based Federated Learning Framework to Defend Against Poisoning Attacks },'' in \emph{Proceedings of the IEEE Symposium on Security and Privacy}, 2024, pp. 2572--2590.

\bibitem{deepsight}
P.~Rieger, T.~Nguyen, M.~Miettinen, and A.-R. Sadeghi, ``{DeepSight}: Mitigating backdoor attacks in federated learning through deep model inspection,'' in \emph{Proceedings of the Network and Distributed Systems Security Symposium}, 2022.

\bibitem{Auror}
S.~Shen, S.~Tople, and P.~Saxena, ``Auror: defending against poisoning attacks in collaborative deep learning systems,'' in \emph{Proceedings of the Annual Conference on Computer Security Applications}, 2016, p. 508–519.

\bibitem{foolsgold}
C.~Fung, C.~Yoon, and I.~Beschastnikh, ``The limitations of federated learning in sybil settings,'' in \emph{Proceedings of the International Symposium on Research in Attacks, Intrusions and Defenses}, Oct. 2020, pp. 301--316.

\bibitem{flame}
T.~D. Nguyen, P.~Rieger, H.~Chen, H.~Yalame, H.~M{\"o}llering, H.~Fereidooni, S.~Marchal, M.~Miettinen, A.~Mirhoseini, S.~Zeitouni, F.~Koushanfar, A.-R. Sadeghi, and T.~Schneider, ``{FLAME}: Taming backdoors in federated learning,'' in \emph{Proceedings of the USENIX Security Symposium}, 2022, pp. 1415--1432.

\bibitem{crowdguard}
P.~Rieger, T.~Krauß, M.~Miettinen, A.~Dmitrienko, and A.-S. Sadeghi, ``Crowdguard: Federated backdoor detection in federated learning,'' in \emph{Proceedings of the Network and Distributed Systems Security Symposium}, 2024.

\bibitem{freqfed}
H.~Fereidooni, A.~Pegoraro, P.~Rieger, A.~Dmitrienko, and A.-R. Sadeghi, ``Freqfed: A frequency analysis-based approach for mitigating poisoning attacks in federated learning,'' in \emph{Proceedings of the Network and Distributed Systems Security Symposium}, 2024.

\bibitem{MESAS}
T.~Krau\ss{} and A.~Dmitrienko, ``{MESAS}: Poisoning defense for federated learning resilient against adaptive attackers,'' in \emph{Proceedings of the ACM Conference on Computer and Communications Security}, 2023, pp. 1526--1540.

\bibitem{BayBFed}
K.~Kumari, P.~Rieger, H.~Fereidooni, M.~Jadliwala, and A.~Sadeghi, ``{BayBFed}: Bayesian backdoor defense for federated learning,'' in \emph{Proceedings of the IEEE Symposium on Security and Privacy}, 2023, pp. 737--754.

\bibitem{FLDetector}
Z.~Zhang, X.~Cao, J.~Jia, and N.~Z. Gong, ``{FLDetector}: Defending federated learning against model poisoning attacks via detecting malicious clients,'' in \emph{Proceedings of the ACM Conference on Knowledge Discovery and Data Mining}, 2022, pp. 2545--2555.

\bibitem{FLTrust}
X.~Cao, M.~Fang, J.~Liu, and N.~Z. Gong, ``{FLTrust}: Byzantine-robust federated learning via trust bootstrapping,'' in \emph{Proceedings of the Network and Distributed Systems Security Symposium}, 2021.

\bibitem{Xie21}
C.~Xie, M.~Chen, P.-Y. Chen, and B.~Li, ``{CRFL}: Certifiably robust federated learning against backdoor attacks,'' in \emph{Proceedings of the International Conference on Machine Learning}, 2021, pp. 11\,372--11\,382.

\bibitem{FedAvg_new}
\BIBentryALTinterwordspacing
H.~McMahan, E.~Moore, D.~Ramage, S.~Hampson, and B.~Arcas, ``Communication-efficient learning of deep networks from decentralized data,'' \emph{CoRR}, vol. abs/1602.05629v4, 2023. [Online]. Available: \url{http://arxiv.org/abs/1602.05629}
\BIBentrySTDinterwordspacing

\bibitem{Konkey2017}
J.~Konečný, B.~McMahan, F.~Yu, P.~Richtarik, A.~Suresh, and D.~Bacon, ``Federated learning: Strategies for improving communication efficiency,'' in \emph{NIPS Workshop on Private Multi-Party Machine Learning}, 2016.

\bibitem{Shejwalkar22}
V.~Shejwalkar, A.~Houmansadr, P.~Kairouz, and D.~Ramage, ``Back to the drawing board: A critical evaluation of poisoning attacks on production federated learning,'' in \emph{Proceedings of the IEEE Symposium on Security and Privacy}, 2022, pp. 1354--1371.

\bibitem{krum}
P.~Blanchard, E.~M. El~Mhamdi, R.~Guerraoui, and J.~Stainer, ``Machine learning with adversaries: Byzantine tolerant gradient descent,'' in \emph{Proceedings of the Advances in Neural Information Processing Systems}, 2017, pp. 118--128.

\bibitem{Xiao12}
H.~Xiao, H.~Xiao, and C.~Eckert, ``Adversarial label flips attack on support vector machines,'' in \emph{Proceedings of the European Conference on Artificial Intelligence}, 2012, p. 870–875.

\bibitem{Krizhevsky09}
\BIBentryALTinterwordspacing
A.~Krizhevsky, V.~Nair, and G.~Hinton, ``Cifar-10 (canadian institute for advanced research),'' 2009. [Online]. Available: \url{http://www.cs.toronto.edu/~kriz/cifar.html}
\BIBentrySTDinterwordspacing

\bibitem{Cohen17}
G.~Cohen, S.~Afshar, J.~Tapson, and A.~van Schaik, ``{EMNIST: Extending MNIST to handwritten letters},'' in \emph{Proceedings of the International Joint Conference on Neural Networks}, 2017, pp. 2921--2926.

\bibitem{He16}
K.~He, X.~Zhang, S.~Ren, and J.~Sun, ``Deep residual learning for image recognition,'' in \emph{Proceedings of the IEEE Conference on Computer Vision and Pattern Recognition}, 2016, pp. 770--778.

\bibitem{Lecun98}
Y.~Lecun, L.~Bottou, Y.~Bengio, and P.~Haffner, ``Gradient-based learning applied to document recognition,'' \emph{Proceedings of the IEEE}, vol.~86, no.~11, pp. 2278--2324, 1998.

\bibitem{Kusetogullari19}
H.~Kusetogullari, A.~Yavariabdi, A.~Cheddad, H.~Grahn, and J.~Hall, ``Ardis: A swedish historical handwritten digit dataset,'' \emph{Neural Computing and Applications}, vol.~32, pp. 16\,505--16\,518, 2020.

\bibitem{HDBSan}
L.~McInnes and J.~Healy, ``Accelerated hierarchical density based clustering,'' in \emph{IEEE International Conference on Data Mining Workshops}, 2017, pp. 33--42.

\bibitem{Baruch19}
M.~Baruch, G.~Baruch, and Y.~Goldberg, ``A little is enough: Circumventing defenses for distributed learning,'' in \emph{Proceedings of the International Conference on Neural Information Processing Systems}, 2019, pp. 8635--8645.

\bibitem{Kraub24}
T.~Krauß, J.~König, A.~Dmitrienko, and C.~Kanzow, ``Automatic adversarial adaption for stealthy poisoning attacks in federated learning,'' in \emph{Proceedings of the Network and Distributed System Security Symposium}, 2024.

\bibitem{Pillutla_defesa20}
K.~Pillutla, S.~M. Kakade, and Z.~Harchaoui, ``Robust aggregation for federated learning,'' \emph{IEEE Transactions on Signal Processing}, vol.~70, pp. 1142--1154, 2022.

\bibitem{Yin_defesa18}
D.~Yin, Y.~Chen, R.~Kannan, and P.~Bartlett, ``Byzantine-robust distributed learning: Towards optimal statistical rates,'' in \emph{Proceedings of the International Conference on Machine Learning}, 2018, pp. 5650--5659.

\bibitem{Guo21}
Y.~Guo, Q.~Wang, T.~Ji, X.~Wang, and P.~Li, ``Resisting distributed backdoor attacks in federated learning: A dynamic norm clipping approach,'' in \emph{Proceedings of the IEEE International Conference on Big Data}, 2021, pp. 1172--1182.

\bibitem{Andreina_defesa21}
S.~Andreina, G.~A. Marson, H.~Möllering, and G.~Karame, ``{BaFFLe}: Backdoor detection via feedback-based federated learning,'' in \emph{Proceedings of the IEEE International Conference on Distributed Computing Systems}, 2021, pp. 852--863.

\bibitem{Brendan18}
\BIBentryALTinterwordspacing
H.~McMahan, D.~Ramage, K.~Talwar, and L.~Zhang, ``Learning differentially private recurrent language models,'' in \emph{Proceedings of the International Conference on Learning Representations}, 2018. [Online]. Available: \url{https://openreview.net/forum?id=BJ0hF1Z0b}
\BIBentrySTDinterwordspacing

\bibitem{deep_learning_diff_privacy}
M.~Abadi, A.~Chu, I.~Goodfellow, H.~B. McMahan, I.~Mironov, K.~Talwar, and L.~Zhang, ``Deep learning with differential privacy,'' in \emph{Proceedings of the ACM Conference on Computer and Communications Security}, 2016, pp. 308--318.

\bibitem{Wei20}
K.~Wei, J.~Li, M.~Ding, C.~Ma, H.~H. Yang, F.~Farokhi, S.~Jin, T.~Quek, and H.~Poor, ``Federated learning with differential privacy: Algorithms and performance analysis,'' \emph{IEEE Transactions on Information Forensics and Security}, pp. 3454--3469, 2020.

\end{thebibliography}

\end{document}